\pdfoutput=1

\documentclass{aa}

\usepackage{graphicx}
\usepackage[varg]{txfonts}
\usepackage{amsmath}
\usepackage{verbatim}
\usepackage{latexsym}
\usepackage{multirow}
\usepackage{fixltx2e}
\usepackage{color}

\newcommand{\ind}[1]{_{\mathrm{#1}}}
\newcommand{\de}{\mbox{d}}

\begin{document}


\title{Measuring Planetary Atmospheric Dynamics with Doppler Spectroscopy}


\author{Patrick Gaulme\inst{1,2,3}
\and 
Fran\c cois-Xavier Schmider\inst{4} 
\and 
Ivan Gon\c{c}alves\inst{4}
}

\institute{Max-Planck-Institut f\"ur Sonnensystemforschung, Justus-von-Liebig-Weg 3, 37077, G\"ottingen, Germany \email{gaulme@mps.mpg.de}
\and 
Department of Astronomy, New Mexico State University, P.O. Box 30001, MSC 4500, Las Cruces, NM 88003-8001, USA 
\and
Physics Department, New Mexico Institute of Mining and Technology, 801 Leroy Place, Socorro, NM 87801, USA 
\and
Laboratoire Lagrange, Universit\'e de Nice Sophia Antipolis, UMR 7293, Observatoire de la C\^ote d'Azur (OCA), Nice, France 
}

\date{Received <date> /
Accepted <date>}

\abstract{
Doppler imaging spectroscopy is the most reliable way to directly measure wind speeds of planetary atmospheres of the Solar system. However, most knowledge about atmospheric dynamics has been obtained with cloud-tracking technique, which consists of tracking visible features from images taken at different dates. Doppler imaging is as challenging -- motions can be less than 100 m s$^{-1}$ -- as appealing because it measures the speed of cloud particles instead of large cloud structures. Significant difference is expected in case of atmospheric waves interfering with cloud structures. The purpose of this paper is to provide a theoretical basis for conducting accurate Doppler measurements of planetary atmospheres, especially from the ground with reflected solar absorption lines. We focus on three aspects which lead to significant biases. Firstly, we fully review the \textit{Young} effect, which is an artificial radial velocity field caused by the solar rotation that mimics a retrograde planetary rotation. Secondly, we extensively study the impact of atmospheric seeing and show that it modifies the apparent location of the planet in the sky whenever the planet is not observed at full phase (opposition). Besides, the seeing convolves regions of variable radial velocity and photometry, which biases radial-velocity measurements, by reducing the apparent amplitude of atmospheric motions. Finally, we propose a method to interpret data, i.e., how to retrieve zonal, meridional, vertical, and subsolar-to-antisolar circulation from radial velocity maps, by optimizing the signal to noise ratio.}

\maketitle


\section{Introduction}
\label{sect_intro}
For dense atmospheres in the solar system, wind measurements mostly come from the cloud tracking technique. The method consists of following cloud features at specific wavelengths taken on image pairs obtained at different times, to retrieve the wind speed before cloud structures evolve or disappear \citep{Sanchez_Lavega_2008, Peralta_2008, Moissl_2009, Peralta_2012}. This method has also provided results with spacecrafts that flew-by or orbited planets \citep[e.g.][]{Choi_2007}. Cloud tracking indicates the motion of large cloud structures (limited by spatial resolution), which is rather an indication of the speed of iso-pressure regions, than the speed of the actual cloud particles. A complementary solution to access direct wind speed measurement is Doppler spectrometry, because it measures the actual speed of cloud particles.
So far, most observational efforts with Doppler measurements of dense planetary atmospheres have been dedicated to Venus and Jupiter, for two distinct reasons. 

Doppler spectrometry was envisioned for Venus to understand its atmospheric circulation as it is a mostly featureless planet in the visible domain. First attempts based on visible spectrometry were performed in the 1970s, but did not lead to robust results \citep{Traub_Carleton_1975, Young_1979}. In 2007, a significant international effort was organized to support the atmospheric observations of Venus by ESA mission Venus Express (VEx) \citep{Lellouch_Witasse_2008}. The objective was to measure the atmospheric circulation using different spectral ranges, to probe different altitudes in the Venus mesosphere. Significant results on the upper mesospheric dynamics were obtained using mid-infrared heterodyne spectroscopy \citep{Sornig_2008, Sornig_2012} and millimeter and submillimeter wave spectroscopy \citep{Clancy_2008,Clancy_2012,Lellouch_2008,Moullet_2012} but Doppler spectroscopy is more challenging at shorter wavelength. Visible observations of solar Fraunhofer lines scattered by Venus clouds were performed by \citet{Widemann_2007, Widemann_2008, Gabsi_2008, Gaulme_2008, Machado_2012, Machado_2014, Machado_2017}. Regarding other planets, similar measurements were performed in the mm/sub-mm domain for Mars \citep{Lellouch_1991, Moreno_2009} and Titan \citep{Moreno_2005}, as well as in the infrared with the 10-$\mu$m heterodyne observations of Titan by \citet{Kostiuk_2001,Kostiuk_2005,Kostiuk_2006,Kostiuk_2010}. 

The use of Doppler spectrometry for giant planets, Jupiter in particular, was inspired by its success with helioseismology to probe their deep internal structures \citep[e.g.,][]{Appourchaux_Grundahl_2013}. The principle relies on the detection of global oscillation modes, whose properties are function of the internal density profile. Giant planets being mostly fluid and convective, their seismology is much closer to that of solar-like stars than that of terrestrial planets. The basic principle relies on monitoring the position of a spectral line that probes an atmospheric level where the amplitude of acoustic modes is maximum. So far, resonant cell \citep{Schmider_1991,Cacciani_2001} and Fourier transform spectrometry \citep[FTS, e.g.,][]{Mosser_1993,Schmider_2007,Goncalves_2016} have been considered. Two dedicated instruments were subsequently designed: the SYMPA instrument was an FTS based on a Mach-Zehnder interferometer \citep{Schmider_2007} which has provided so far the clearest observational evidence of Jupiter's oscillations \citep{Gaulme_2011}. Since then, the new instrument JOVIAL/JIVE, inherited from SYMPA, has been developed and tested to perform both atmospheric dynamics of dense atmospheres and seismic measurements of Jupiter and Saturn \citep[][and submitted]{Goncalves_2016}.

In this paper, we first study the influence of two possibles sources of biases regarding Doppler spectroscopic measurements of planetary atmospheres. On one hand, we review and extensively revise the impact of the solar rotation on radial velocity measurements, which was originally studied by \citet{Young_1975} in the case of Venus (Sect. \ref{sect_young}). On the other hand, we consider the impact of atmospheric seeing on both positioning on the planet and on radial velocity maps (Sect. \ref{sect_seeing}). In a third part, we focus on how to retrieve zonal and meridional components of atmospheric circulation from radial velocity maps (Sect. \ref{sect_model}).

\section{Effect of the apparent diameter of the Sun}
\label{sect_young}
\subsection{The Young effect}
\citet{Young_1975} was the first to point out that the apparent diameter of the Sun as seen from Venus causes an artificial radial velocity field on it. Indeed, rays from different parts of the Sun, which display different radial velocities, reach the planet with (slightly) different incidence angles. 
Regions of the Sun that are closer to Venus' horizon contribute less to the reflected solar spectrum than regions closer to zenith. Thus, the radial velocity integrated over the whole solar disk is not zero in a given point of Venus. In other words, even if Venus would not be rotating, we would still measure a Doppler shift near the Venus terminator, and that Doppler shift would mimic a retrograde rotation because the solar rotation is prograde. 

Despite its pioneering aspect, \citet{Young_1975}'s analytical description of the artificial Doppler shift $\Delta V\ind{Y}$ on Venus was limited to the equator. Out of the equator, the expression $\Delta V\ind{Y} \approx 3.2 \tan \gamma$, where $\gamma$ is the solar-zenith angle, is no more valid.  Nonetheless, most authors who hitherto worked with Doppler spectroscopic data of Venus in the visible used Young's formulation, without noticing it was valid at the equator only,  and certainly got biases in their results and conclusions \citep{Widemann_2007, Widemann_2008, Gabsi_2008, Machado_2012, Machado_2014, Machado_2017}. We propose a revised analytical description of the ``Young'' effect with the same approximations used by \citet{Young_1975}, as well as a more comprehensive numerical simulation. 

\begin{figure}[t!]
\center
\includegraphics[width=7cm]{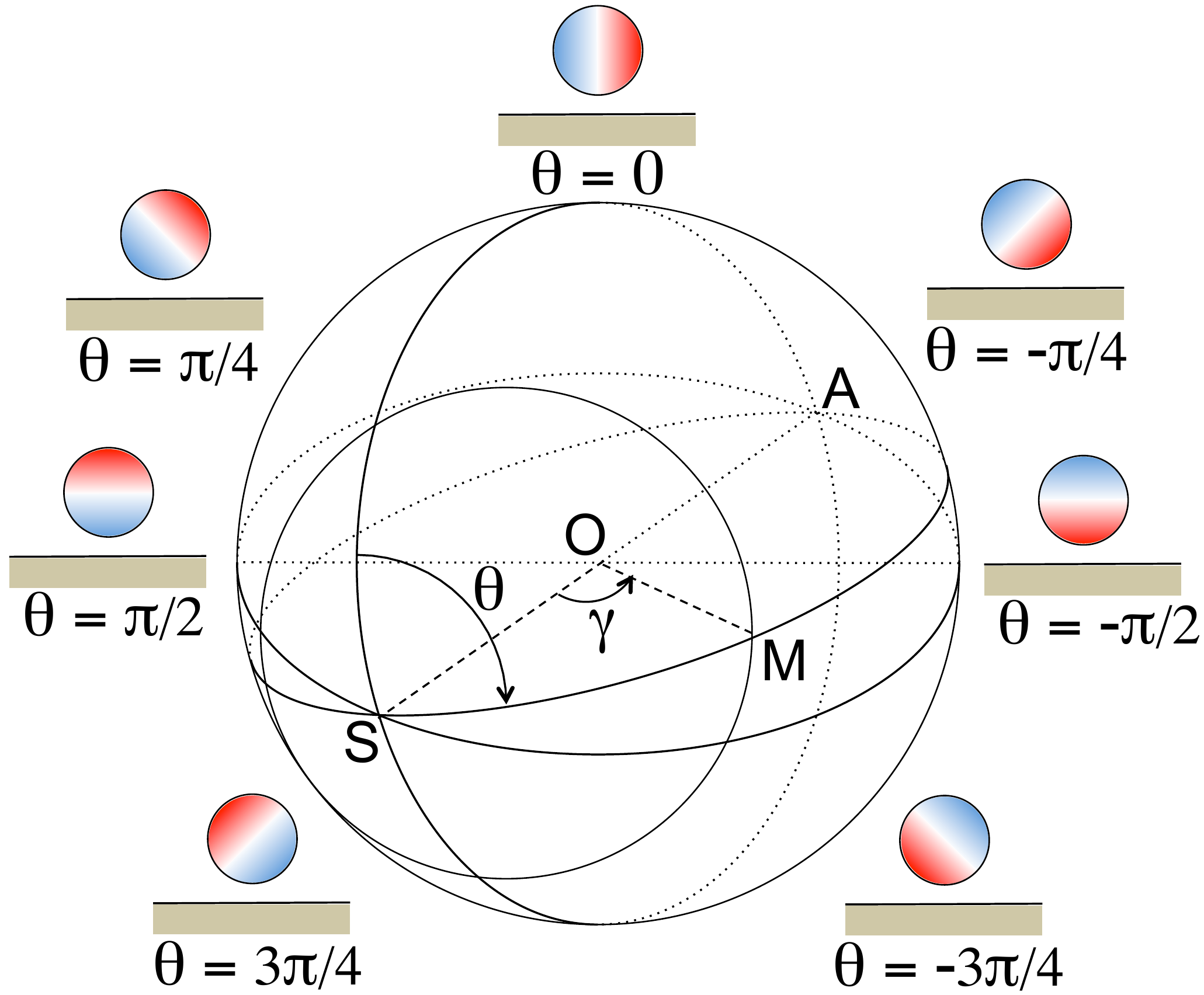} 
\caption{Reference frame used to express the \textit{Young} effect. The points S and A are the sub-solar and anti-solar points respectively, O is the center of the coordinates and M a given point on the planet. In this frame, the angles of incidence $\gamma$ and horizon inclination $\theta$ are uniform along parallels and meridians respectively. The sub-pictures represent the radial velocity map of the Sun a seen with $\theta = [0, -\pi/4, -\pi/2]$. The artificial radial velocity is not observed along the meridian $\theta=0$, while it is maximum along $\theta=\pm\pi/2$.}
\label{fig_gamma_theta}
\end{figure}
\subsection{Mathematical description}

In this section, we focus on Venus, as this effect is almost negligible for external planets, even though our analysis is valid for any planet of the solar system. We do not select Mercury as the test case, as such a kind of measurements is useful only to monitor dynamics of optically thick atmospheres and not solid surfaces covered by fixed contrasted features. To express the artificial Doppler shift $\Delta V\ind{Y}$, known as \textit{Young} effect, we assume that the uppermost cloud layer of the considered planet is a flat surface.
Within this assumption, at a given point on Venus, the intensity of a ray coming from the Sun is proportional to the cosine of the incidence angle $\gamma$.

Two factors drive the \textit{Young} effect, which were both mentioned by Young himself but only the first one was implemented. Firstly, the variation of incidence angle from different parts of the Sun results in averaging regions with different radial velocities and intensities. Besides, the center part of the Sun appears brighter than the edge -- limb darkening effect -- which slightly counterbalances the artificial Doppler signal, by giving more importance to regions with low radial velocities. Secondly, the inclination of the solar rotation axis with respect to the horizon modulates the \textit{Young} effect. For example, near Venus poles, where the solar rotation axis is perpendicular to the horizon, both the blue and red sides of the Sun reach the surface in a symmetrical fashion and average out. To the contrary on Venus morning terminator, the blue side of Venus rises first and the artificial Doppler signal is maximum (Fig. \ref{fig_gamma_theta}). This second factor was not taken into account in \citet{Young_1975}'s equation, because he focused on describing the amplitude of the phenomenon along the equator, where it is maximum. 

In other words, the parameters that drive the \textit{Young} effect are: 1) the angle of solar incidence $\gamma$ (solar-zenith angle); 2) the inclination $\theta$ of the solar rotation axis with respect to the local horizon. Let us consider the coordinate system that connects the sub-solar to the anti-solar points of Venus (Fig. \ref{fig_gamma_theta}). The incidence angle $\gamma$ is uniform along parallels of this system of coordinates, while the tilt of the solar spin axis with respect to the horizon $\theta$ is uniform along the meridians. In a given point on Venus of solar incidence $\gamma$ and rotation axis inclination $\theta$, the artificial Doppler effect is the integral of the radial velocity map weighted by the differential photometry: 

{\setlength{\mathindent}{0cm}
\small
\begin{equation}
\Delta V\ind{Y}(\gamma,\theta) = \frac{\int_{-\rho_\odot}^{\rho_\odot} \int_{-\sqrt{\rho_\odot^2 - y^2}}^{\sqrt{\rho_\odot^2 - y^2}}  I\ind{LD}(x,y) \cos(\gamma-y) \Delta V_\odot(x,y,\theta)\ \mathrm{d}x\ \mathrm{d}y}{\int_{-\rho_\odot}^{\rho_\odot} \int_{-\sqrt{\rho_\odot^2 - y^2}}^{\sqrt{\rho_\odot^2 - y^2}}  I\ind{LD}(x,y) \cos(\gamma-y)\ \mathrm{d}x\ \mathrm{d}y}
\label{eq_integrale}
\end{equation}}

\noindent where $\rho_\odot$ is the solar apparent diameter as seen from Venus, $I\ind{LD}(x,y)$  its limb-darkening law,  and $\Delta V_\odot(x,y,\theta)$ is the radial velocity field of the solar surface, expressed in the frame $(x,y)$ centered on the solar disk (Fig. \ref{fig_solar_frame_photo}).

\subsection{Approximated analytical solution}
\begin{figure}[t!]
\center
\includegraphics[width=7cm]{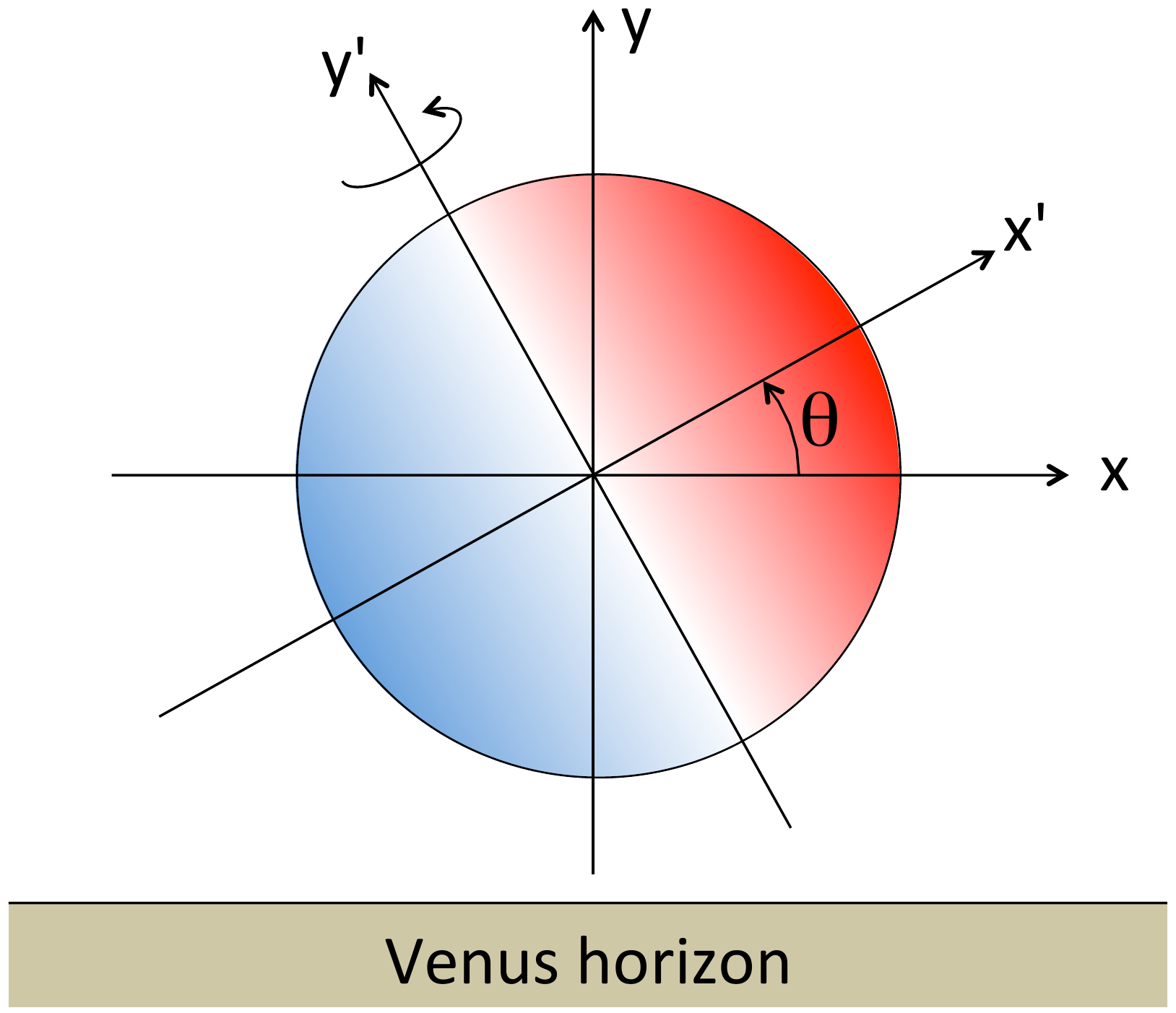}  
\caption{Schematic view of the radial-velocity map of the Sun as seen from Venus. The solar rotation axis is tilted by the angle $\theta$. The system of coordinates $(x,y)$ is the frame in which we compute the integral in Eq. \ref{eq_integrale}, while $(x',y')$ is the system in which the solar rotation is described.  }
\label{fig_solar_frame_RV}
\end{figure}

In this section, we demonstrate the analytical expression of the \textit{Young} effect, by assuming the same approximations that were used by \citet{Young_1975}. The first approximation consists of considering the Sun as a solid rotator, i.e., with uniform rotation period as function of latitude, which is actually quite far from the actual Sun, for which differential is significant \citep[24 days at equator and 38 in polar regions, according to][]{Snodgrass_Ulrich_1990}. The second approximation consists of neglecting any solar limb-darkening, which is not accurate either \citep[e.g.,][]{Allen_1973}. The third approximation consists of considering Venus at top-cloud level as a Lambert scattering surface, i.e. iso-scattering by surface element. 
Notations for angles are described in Fig. \ref{fig_solar_frame_RV}. 

The radial velocity map of a solid rotator as seen from its equatorial plane is a linear function of the coordinates on the considered rotator. Let us consider the $(x',y')$ system centered on the Sun and tilted accordingly to its rotation axis (Fig. \ref{fig_solar_frame_RV}), the radial velocity field $\Delta V_\odot(x',y')$ of the solar surface is: 
\begin{equation}
\Delta V_\odot(x',y') = \frac{V\ind{eq}}{\rho_\odot} x', 
\end{equation}
where $V\ind{eq}$ si the solar equatorial velocity, and $x',y'$, and $\rho_\odot$ are expressed in arcsec. By considering that the Sun is tilted by the angle $\theta$, it re-expresses in the local system $(x,y)$, in which $x$ is parallel to the horizon:
\begin{equation}
\Delta V_\odot(x,y,\theta) = \frac{V\ind{eq}}{\rho_\odot} \left(x \cos\theta + y \sin\theta \right) .
\end{equation}

\begin{figure}[t!]
\center
\includegraphics[width=7cm]{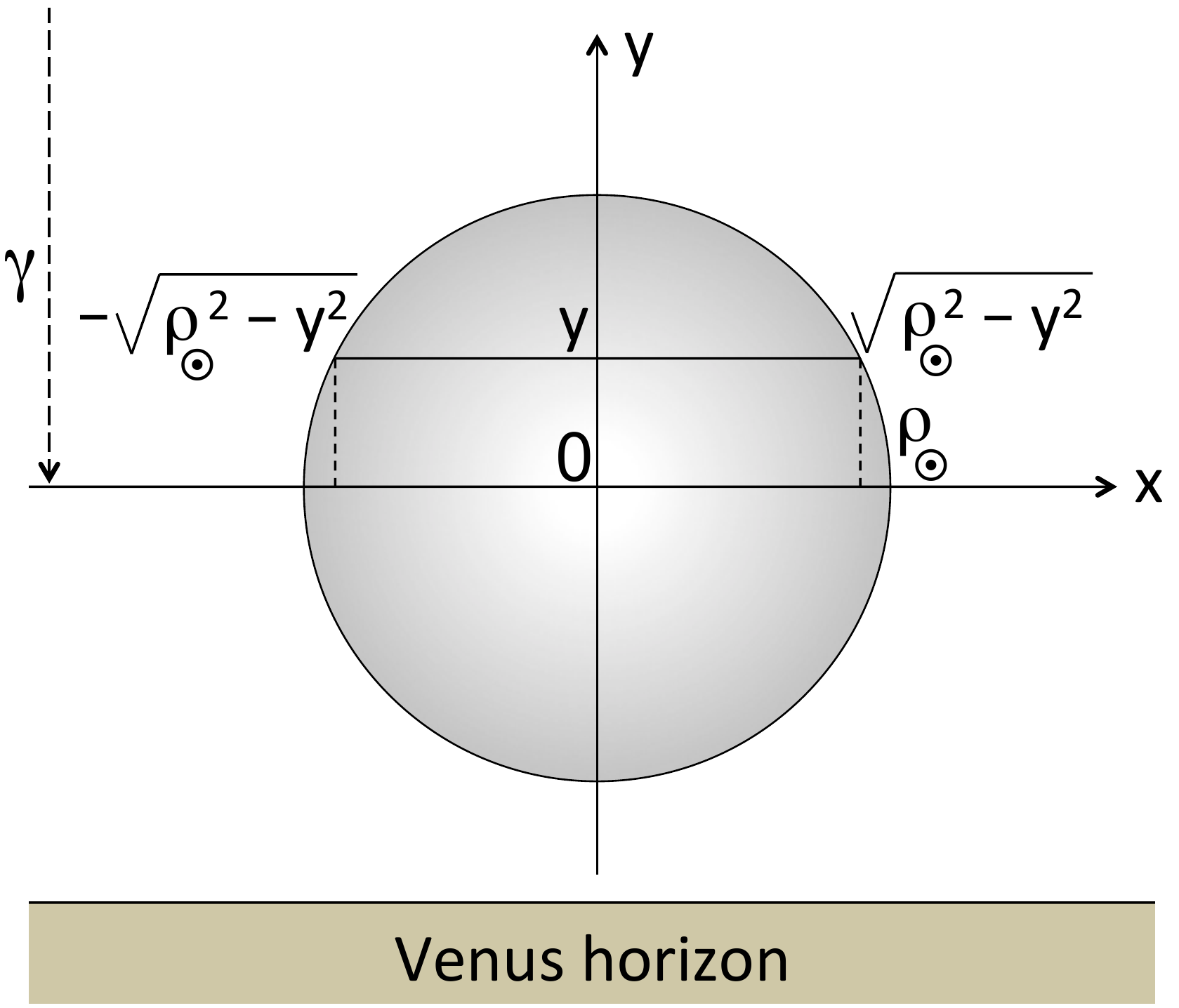}
\caption{Schematic view of the photometric aspect of the Sun as seen from Venus. The angle $\gamma$ is the solar-zenith angle, i.e., the angle between the zenith and the center of the Sun. The system of coordinates $(x,y)$ is centered on the Sun; the $y$-axis points to the zenith. Hence, the ordinate of the solar region closest to the horizon is $\gamma-y$. All variables are angles.}
\label{fig_solar_frame_photo}
\end{figure}

Since we neglect the solar limb-darkening, Eq. \ref{eq_integrale} gets simplified: 
{\setlength{\mathindent}{0cm}
\small
\begin{eqnarray}
\Delta V\ind{Y}(\gamma,\theta)&=&\frac{V\ind{eq}}{\rho_\odot}\ \frac{\int_{-\rho_\odot}^{\rho_\odot} \int_{-\sqrt{\rho_\odot^2 - y^2}}^{\sqrt{\rho_\odot^2 - y^2}}  \cos(\gamma-y)\ \left(x \cos\theta + y \sin\theta \right)\ \mathrm{d}x\ \mathrm{d}y}{\int_{-\rho_\odot}^{\rho_\odot} \int_{-\sqrt{\rho_\odot^2 - y^2}}^{\sqrt{\rho_\odot^2 - y^2}} \cos(\gamma-y)\ \mathrm{d}x\ \mathrm{d}y}\\
&=&\frac{V\ind{eq}}{\rho_\odot}\ \sin\theta\ \frac{\int_{-\rho_\odot}^{\rho_\odot}\ y \sqrt{\rho_\odot^2-y^2}\ \cos(\gamma-y)\ \mathrm{d}y}{\int_{-\rho_\odot}^{\rho_\odot}\ \sqrt{\rho_\odot^2-y^2}\ \cos(\gamma-y)\ \mathrm{d}y}
\label{eq_integrale_simple}
\end{eqnarray}}

By assuming that $y\ll\gamma$, which is true except for the sub-solar area, we can develop $\cos(\gamma-y)$ into $\cos\gamma + y\sin\gamma$ and analytically solve the integral:
\begin{equation}
\Delta V\ind{Y}(\gamma,\theta) =  \frac{V\ind{eq}}{\rho_\odot}\ \sin\theta\ \frac{\sin\gamma}{\cos\gamma}\ \frac{\int_{-\rho_\odot}^{\rho_\odot}\ y^2 \sqrt{\rho_\odot^2-y^2}\  \mathrm{d}y}{\int_{-\rho_\odot}^{\rho_\odot}\ \sqrt{\rho_\odot^2-y^2}\ \mathrm{d}y}
\end{equation}
The integration of the latter expression leads to a revised version of the \textit{Young} effect: 
\begin{equation}
\Delta V\ind{Y}(\gamma,\theta) =  \frac{V\ind{eq}\ \rho_\odot}{4}\ \sin\theta\ \tan\gamma 
\label{eq_young_analytical_simple}
\end{equation}
By assuming a solar equatorial velocity of 2 km s$^{-1}$, and by considering that the apparent solar radius as seen from Venus is $6.43\, 10^{-3}$ rad,  $V\ind{eq}\ \rho_\odot/4 \approx3.2$ m s$^{-1}$. Our expression differs from Young's only by the introduction of the $\sin\theta$ dependence. The numerical simulation that is described in the next section allows us to adjust the amplitude factor -- here 3.2 m s$^{-1}$ -- when we take into account the solar limb darkening and differential rotation. 

As pointed out by previous authors \citep[e.g.,][]{Machado_2012}, the expression of the \textit{Young} effect diverges at the terminator, where $\gamma = \pi/2$. This is because the above expression is a good approximation as long as the Sun is entirely visible from a given point on Venus, i.e., $\gamma < \pi/2 - \rho_\odot$. Once the Sun is partially below the horizon, Eq. \ref{eq_young_analytical_simple} is no more valid, and Eq. \ref{eq_integrale_simple} must be integrated from the horizon line (ordinate $y = \gamma -\pi/2$) to $\rho_\odot$ (see Fig. \ref{fig_solar_frame_photo}). In that case, the \textit{Young} effect near \textit{terminator} becomes: 
\begin{equation}
\displaystyle
\Delta V\ind{Y}(\gamma,\theta) =   \frac{V\ind{eq}\ \rho_\odot}{4} \ \sin{\theta}\ \frac{\Phi\sin\gamma + \mathrm{X}\cos\gamma}{\displaystyle\frac{\rho_\odot^2}{4} \ \mathrm{X}\sin\gamma + \Psi\cos\gamma}
 \label{eq_young_analytical_terminator}
\end{equation}
where
\begin{equation}
\begin{split}
\Phi(\gamma) =\ & \frac{\pi}{2}\ - \ \sin^{-1}\left({\frac{ \gamma - \frac{\pi}{2}}{\rho_\odot}}\right)\ - \ \frac{ \gamma - \frac{\pi}{2}}{\rho_\odot^2} \left[\rho_\odot^2 -  \left(\gamma - \frac{\pi}{2}\right)^2\right]^{\frac{1}{2}}  + \\
&  2\ \frac{ \gamma - \frac{\pi}{2}}{\rho_\odot^4} \left[\rho_\odot^2 -  \left(\gamma - \frac{\pi}{2}\right)^2\right]^{\frac{3}{2}}\
\end{split}
\end{equation}
\begin{equation}
\mathrm{X}(\gamma) =  \frac{8}{3}\ \frac{1}{\rho_\odot^4} \left[\rho_\odot^2 -  \left(\gamma - \frac{\pi}{2}\right)^2\right]^{\frac{3}{2}}\
\end{equation}
\begin{equation}
\Psi(\gamma) =\  \frac{\pi}{2} \ - \ \sin^{-1}\left({\frac{ \gamma - \frac{\pi}{2}}{\rho_\odot}}\right) \ - \ \frac{ \gamma - \frac{\pi}{2}}{\rho_\odot^2} \left[\rho_\odot^2 -  \left(\gamma - \frac{\pi}{2}\right)^2\right]^{\frac{1}{2}}  
\end{equation}
which is valid as long as $\gamma \in \left[\frac{\pi}{2} - \rho_\odot, \frac{\pi}{2}  + \rho_\odot\right]$. Note that we include the expression of the \textit{Young} effect in this configuration for completeness only. Actually, it does not make much sense from a physical point of view since other effects interfere, as the atmospheric refraction. In addition, as we indicate in Sect. \ref{sect_seeing_rv} (Fig. \ref{fig_young_simu_seeing}), the \textit{Young} effect is drastically diminished by the bias on radial velocities caused by instrumental and/or atmospheric Point Spread Function (PSF).

\begin{figure}[t!]
\center
\includegraphics[width=9cm]{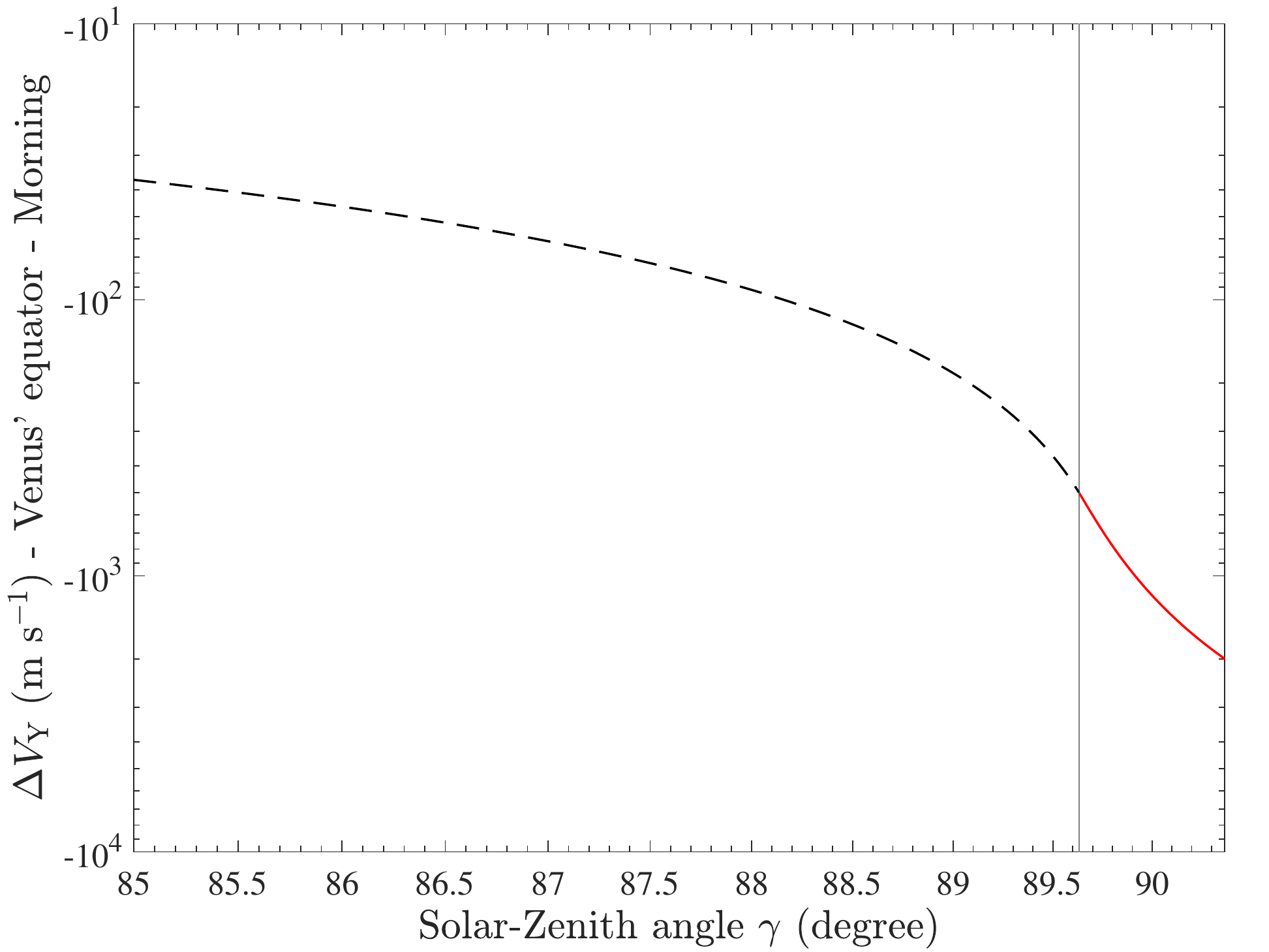}  
\caption{The Young effect computed along Venus's equator on the morning hemisphere for solar-zenith angles $\gamma = [85\pi/180,\pi/2 + \rho_\odot]$. The black dashed line indicates the Young effect in the case the Sun is above the horizon (Eq. \ref{eq_young_analytical_simple}). The plain red line corresponds with the case of the rising Sun (Eq. \ref{eq_young_analytical_terminator}). The vertical line indicates the first contact of the Sun with the horizon. The minimum value of $\Delta V\ind{Y}$ is $-2,000$ m s$^{-1}$ corresponds with the Sun's equatorial velocity $V\ind{eq}$ used for the simulation.}
\label{fig_young_simu_carre}
\end{figure}

\subsection{Numerical simulations}
It is impossible to analytically integrate Eq. (\ref{eq_integrale}) without significant approximations, such as considering $y\ll\gamma$. We opt for a numerical simulation based on the observational knowledge of the Sun's limb darkening and differential rotation. 
The simulations are built into two steps. First, we make a Venus 3-dimension model that we project into a 2-dimension frame, on which the angles $\gamma$ and $\theta$ are known for each point of Venus surface (see Appendix \ref{sect_3D_ell}). Secondly, for each point on Venus, we simulate the rotating Sun and integrate Eq. (\ref{eq_integrale}).

From simple trigonometric considerations, the two angles $\gamma$ and $\theta$ can be expressed in the planet's coordinate system:
\begin{eqnarray}
\gamma &=& \cos^{-1} \left[\cos(\lambda - \lambda\ind{S}) \cos(\phi - \phi\ind{S})\right]\\
\theta &=& \tan^{-1} \left[-\frac{\sin(\phi - \phi\ind{S})}{\tan(\lambda - \lambda\ind{S})}\right]
\end{eqnarray}
where $\lambda$ and $\phi$ are the latitude and longitude of the planet, and $\lambda\ind{S}$ and $\phi\ind{S}$ are the coordinates of the sub-solar point in this frame. Latitudes run from 0 at the equator to $\pi/2$ at the poles in both hemispheres, and longitudes run from $-\pi/2$ at the evening terminator to $\pi/2$ at the morning terminator. 

In each position on Venus we consider both the solar photometric map, i.e. a disk dominated by the limb darknening law weighted by $\cos(\gamma-y)$, and the radial velocity map, which includes the solar differential rotation and the inclination $\theta$ with respect to the local horizon. According to \citet{Allen_1973}, the limb darkening law can be described as a quadratic law:
\begin{equation}
I(\mu) = I(0)\ \left[1 - u_2(1-\mu) - v_2(1-\mu^2)\right],
\label{eq_LD_1}
\end{equation}
where $\mu = \cos\gamma_\odot$, and where $\gamma_\odot$ is the angle between the normal to the solar surface and the direction pointing at the observer. A point located at a radius $r$ from the solar center is function of the $\gamma_\odot$ angle as $r = R \cos\gamma_\odot$. Hence, $\mu = \cos \left(\sin^{-1}(r/R)\right)$, i.e., $\mu = \sqrt{1 - (r/R)^2}$. Therefore, Eq. (\ref{eq_LD_1}) can be expressed in the $(x,y)$ coordinate system:
\begin{equation}
I\ind{LD}(x,y) = 1\ -\ u_2\left(1- \sqrt{1 - \frac{x^2 + y^2}{R^2}}\right)\ -\ v_2\ \frac{x^2 + y^2}{R^2}
\label{eq_LD_2}
\end{equation}
where we set $I(0) = 1$. From \citet{Allen_1973}, the solar limb darkening coefficients are $u_2 $ and $v_2$ are given in a table that covers the whole visible range.

\begin{figure}[t!]
\center
\includegraphics[width=9cm]{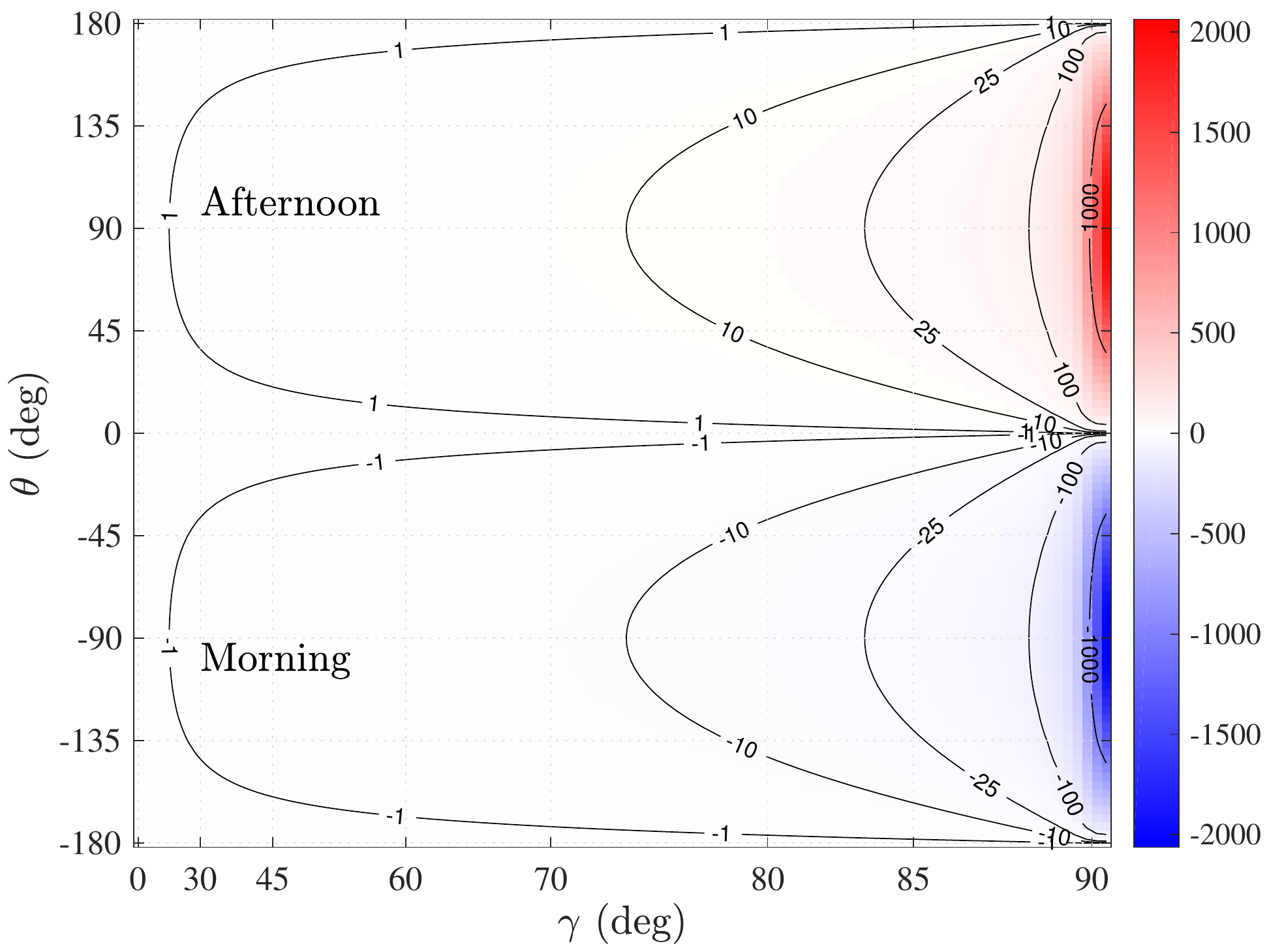}  
\caption{Numerical computation of the \textit{Young} effect for Venus as function of the incidence angle $\gamma$ and the solar rotation axis inclination with respect to the horizon $\theta$, both expressed in degrees. Note the $x$-axis is not linear to enhance the large values of the effect at large incidence angles. Angles $\theta>0$ correspond to the afternoon hemisphere of Venus, and $\theta<0$ to the morning hemisphere. Contours and color scale are in m s$^{-1}$.}
\label{fig_young_simu_carre}
\end{figure}
\begin{figure}[t!]
\center
\includegraphics[width=9cm]{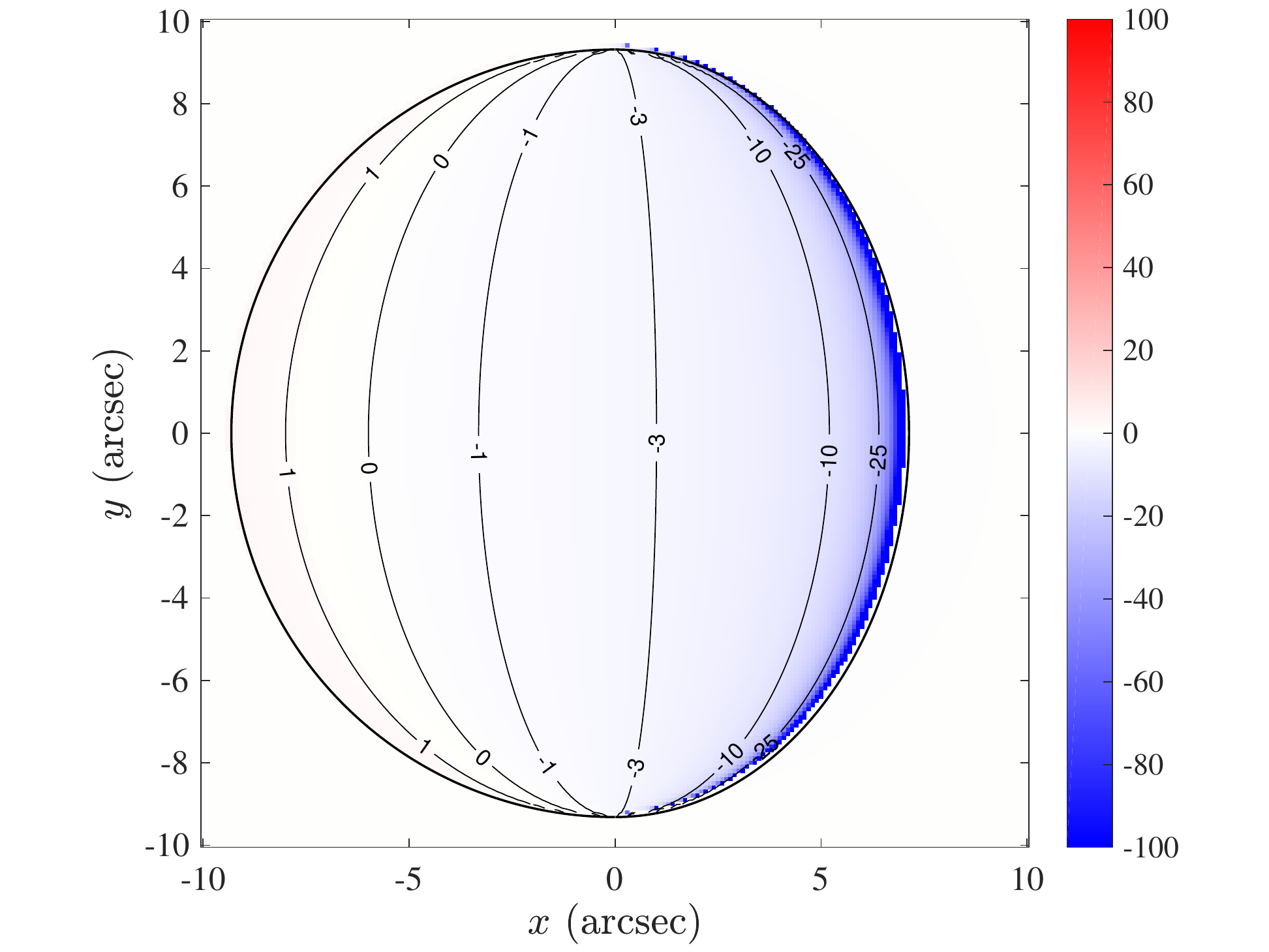}  
\caption{Numerical computation of the \textit{Young} effect for Venus as seen from the Earth with a phase angle (Sun-Venus-Earth) of 40$^\circ$ and a apparent diameter of 11.6 arcsec. The morning hemisphere is mostly visible. The 0$^\circ$ m s$^{-1}$ isovalue corresponds with the meridian crossing the sub-solar point of Venus. Contours and color scale are in m s$^{-1}$.}
\label{fig_young_simu_sphere}
\end{figure}

According to \citet{Snodgrass_Ulrich_1990}, the angular velocity of the solar differential rotation can be described with the function:
\begin{equation}
\Omega_\odot = A + B \sin^2\lambda_\odot + C \sin^4\lambda_\odot
\end{equation}
where $\lambda_\odot$ is the latitude on the Sun, and $A = 2.97\pm0.01$,  $B = -0.48\pm0.04$, and $B = -0.36\pm0.05\ \mu$rad s$^{-1}$. The rotation velocity map is then $R_\odot\,\Omega_\odot\,\cos\lambda_\odot$, and the radial velocity map $\Delta V_\odot$ is the projection of it towards Venus:
\begin{equation}
\Delta V_\odot = R_\odot\ \left(A + B \sin^2\lambda_\odot + C \sin^4\lambda_\odot\right)\ \cos\lambda_\odot\ \cos\phi_\odot,
\end{equation}
where $R_\odot$ is the solar radius in meters, and $\phi_\odot$ the solar longitude as seen from Venus, ranging from 0 to $\pi$ from West to East. The numerical integration of Eq. (\ref{eq_integrale}) is therefore the sum of the solar radial velocity map $\Delta V_\odot$ times the photometric contribution $I\ind{LD} \cos(\gamma-y)$, then normalized by the sum of the photometric term. We simulated the Sun with a $256\times256$ element grid, which is enough in terms of precision. Note that next to the terminator, we take into account that the Sun is partially below the horizon. The maximum value of the \textit{Young} effect corresponds to the configuration where only a thin portion of solar equator is visible, i.e., about 2 km s$^{-1}$. We represent the \textit{Young} effect both for a 2D-map with $\theta$ ranging from -180 to 180$^\circ$ and $\gamma$ from 0 to 90$^\circ$ (Fig. \ref{fig_young_simu_carre}), and also for a example of Venus figure as seen from the Earth. Simulations were done with a limb darkening law corresponding to 550 nm optical wavelength. The two figures highlight the fact that this effect is significant only near the terminator, as it is less than $\pm 10$ m s$^{-1}$ at incidence lower than $\approx75^\circ$, whereas zonal winds are expected to exceed 100 m s$^{-1}$.

\begin{figure}[t!]
\center
\includegraphics[width=9cm]{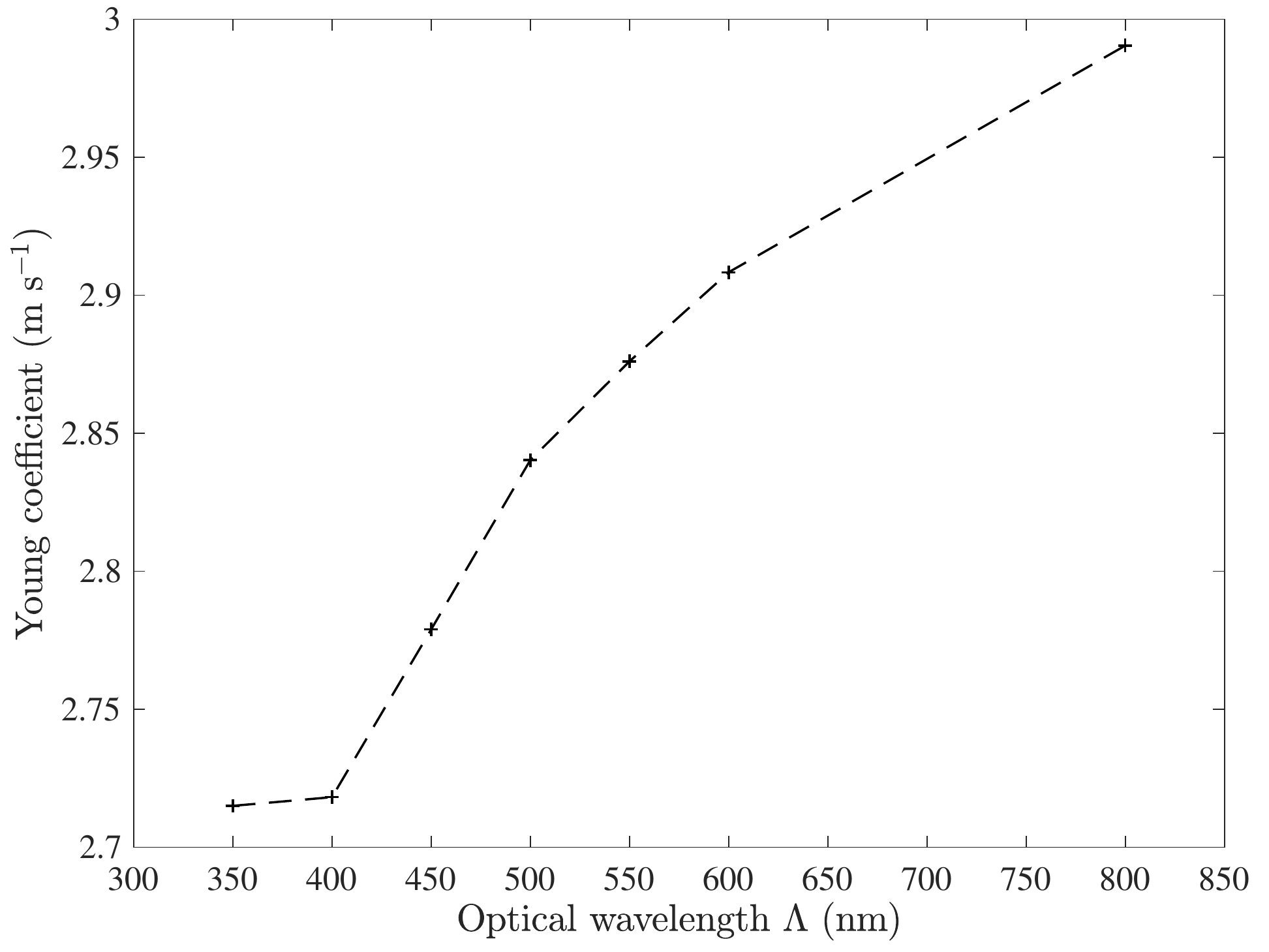}  
\caption{Coefficient of the  \textit{Young} effect for Venus as function of the optical wavelength, from the \citet{Allen_1973} limb darkening laws.}
\label{fig_young_coef}
\end{figure}

To determine the empirical coefficient of the \textit{Young} effect, we computed the ratio of the numerical \textit{Young}-effect map to $\tan\gamma \sin\theta$. We also considered limb-darkening laws that covers the whole visible range (Fig. \ref{fig_young_coef}). The coefficient appears to range from 2.7 to 3 from 350 to 800 nm, which is logically smaller than the 3.2 value, obtained by neglecting the Sun's limb-darkening. We suggest to use the following expression for the \textit{Young} effect: 
\begin{equation}
\Delta V\ind{Y}(\gamma,\theta) =  Y(\Lambda)\ \sin\theta\ \tan\gamma 
\label{eq_young_numerical_coef}
\end{equation}
where $\Lambda$ indicates the optical wavelength, and the coefficient $Y(\Lambda)$ is indicated in Table \ref{tab_young} for the major planets of the Solar system. Even it is useless for measuring atmospheric dynamics, we include Mercury and Mars in Table \ref{tab_young} for completeness and also because detecting the \textit{Young} effect in a solid-rotating body would be a good test for our ability to lead high precise radial velocity measurements. 

Again, near the terminator, the expression of the \textit{Young} effect must take into account that the Sun is partially below the horizon:
\begin{equation}
\displaystyle
\begin{split}
\Delta V\ind{Y}(\gamma,\theta) = &\  Y(\Lambda)\ \sin{\theta}\ \frac{\Phi\sin\gamma + \mathrm{X}\cos\gamma}{\displaystyle\frac{\rho_\odot^2}{4} \ \mathrm{X}\sin\gamma + \Psi\cos\gamma},
 \label{eq_young_numerical_coef_terminator}
\end{split}
\end{equation}
with the same notations as in Eq. \ref{eq_young_analytical_terminator}.

\begin{table}[t!]
\small
 \caption{Coefficient of the  \textit{Young} effect for Venus, Jupiter, Saturn, Uranus, and Neptune, for an optical wavelengths of 350, 550, and 800 nm from the \citet{Allen_1973} limb darkening laws. Orbital semi-major axes are used as Sun-to-planet distances. }
\center
\begin{tabular}{l c c c}\hline
Planet  &	$Y$ (350 nm) &  $Y$ (550 nm)  & $Y$ (800 nm) \\
\hline
Mercury & 5.07	& 	5.37 	& 5.59	\\
Venus & 2.71	& 2.88 	& 2.99\\
Mars  &  1.29    & 1.36      &  1.42    \\
Jupiter &	0.38	& 0.40	& 0.42\\
Saturn/Titan &	0.20	& 0.22	& 0.23\\
Uranus & 	0.10	& 0.10	& 0.11\\
Neptune & 0.07		& 0.07 	& 0.07\\
\hline
\end{tabular}
\label{tab_young}
\end{table}


\section{Impact of atmospheric seeing}
\label{sect_seeing}
\subsection{Biases caused by the point spread function}
Instrumental PSF is another major cause of biases in planetary observations. 
Since we mainly consider earth-based observations in the visible, the PSF is dominated by the atmospheric seeing, so we hereafter refer to the effects of atmospheric seeing instead of instrumental PSF. However, the effects that we describe are also valid for any PSF, even in the case of space-borne measurements, although the biases we identify are much smaller. 

Firstly, seeing convolves the planetary shape and photometry by the PSF, which is usually described as a Gaussian. For external planets at their opposition, the seeing simply blurs the planetary image, which is not a major issue, except for defining the edge. Indeed, it is not alway possible to perfectly know the pixel field-of-view (FOV) of an instrument mounted at a telescope focus, and convert the ephemeris apparent diameter from arcsec to pixels\footnote{This comment about possible inaccurate pixel FOV refers to the case of Venus observations by Gaulme et al. (in prep.) where the MultiRaies spectrometer of the THEMIS solar telescope was used in a specific configuration. The pixel FOV was know to about 5\,\% and no calibration was done to measure it.}. For planets that are out of opposition, especially the inner planets Mercury and Venus, important phase angles (Sun-planet-observer) lead to significant mislocation of the planet because the blurring effect caused by the atmospheric seeing is not the same at limb or terminator edges. Secondly, besides positioning issues, atmospheric seeing blends regions with different photometry \textit{and} radial velocity, which \textit{biases} radial velocity measurements, more importantly towards the planetary edges, where photometric variations are steep. We insist on the term \textit{bias} because atmospheric seeing is often seen as a phenomenon that smooths out radial velocities, while it actually introduces systematic biases in the measurements, which lead to wrong estimates of the wind circulation even with excellent SNR data (see Sect. \ref{sect_seeing_rv}, and Fig. \ref{fig_RV_seeing_cut}).

\subsection{Positioning on the planet}
\label{sect_seeing_pos}
A fundamental step in measuring winds of a planet with a mostly featureless dense atmosphere is to make sure to know where the telescope is slewing. If ephemeris are very accurate for the main solar system bodies, pointing errors of any telescope are typically of about 10 arcsec, which is not enough for an extended object of that kind of apparent size. The use of a guider camera then allows the observer to position the spectrometer's pinhole/slit on the planetary disk. However, this step is not as accurate as one could think. 

There are commonly two ways of determining the location on an extended object as a planet. A first approach is based on the calculation of the photometric centroid of the image, by assuming the background to be negligible. This statement is rather uncertain because the background signal is generally not uniform and scattering is often present around any bright object. This could bias the determination of the photocenter. Moreover, some other factors may bias the photocenter with time. For instance, in the case of Jupiter, visible atmospheric features, moon transits and their cast shadows alter the measured photocenter. 
The second approach consists of  identifying the planetary edge on the image. If the pixel FOV is perfectly known, the observer applies a photometric threshold on the actual image until the portion of the planet above the threshold matches the expected size of the planet on the detector. If the pixel FOV is not perfectly known, a photometric threshold on the image refines it. For example an observer can assume that the planet is all what is above 5\,\% of the maximum intensity, and adjust the pixel FOV accordingly.  Since this is rather arbitrary, we assume that the pixel size can be known (it can be calibrated using double stars or motion applied to the telescope).
In the following, we only consider the second approach. We show that the size of the PSF leads to misestimating the planet's position on the detector, and we determine what correction must be applied. Note that the photometric centroid does not suffer from such a bias related to atmospheric seeing, but as said earlier, it could be affected by other uncalibrated variations.

\begin{figure}[t!]
\center
\includegraphics[width=9cm]{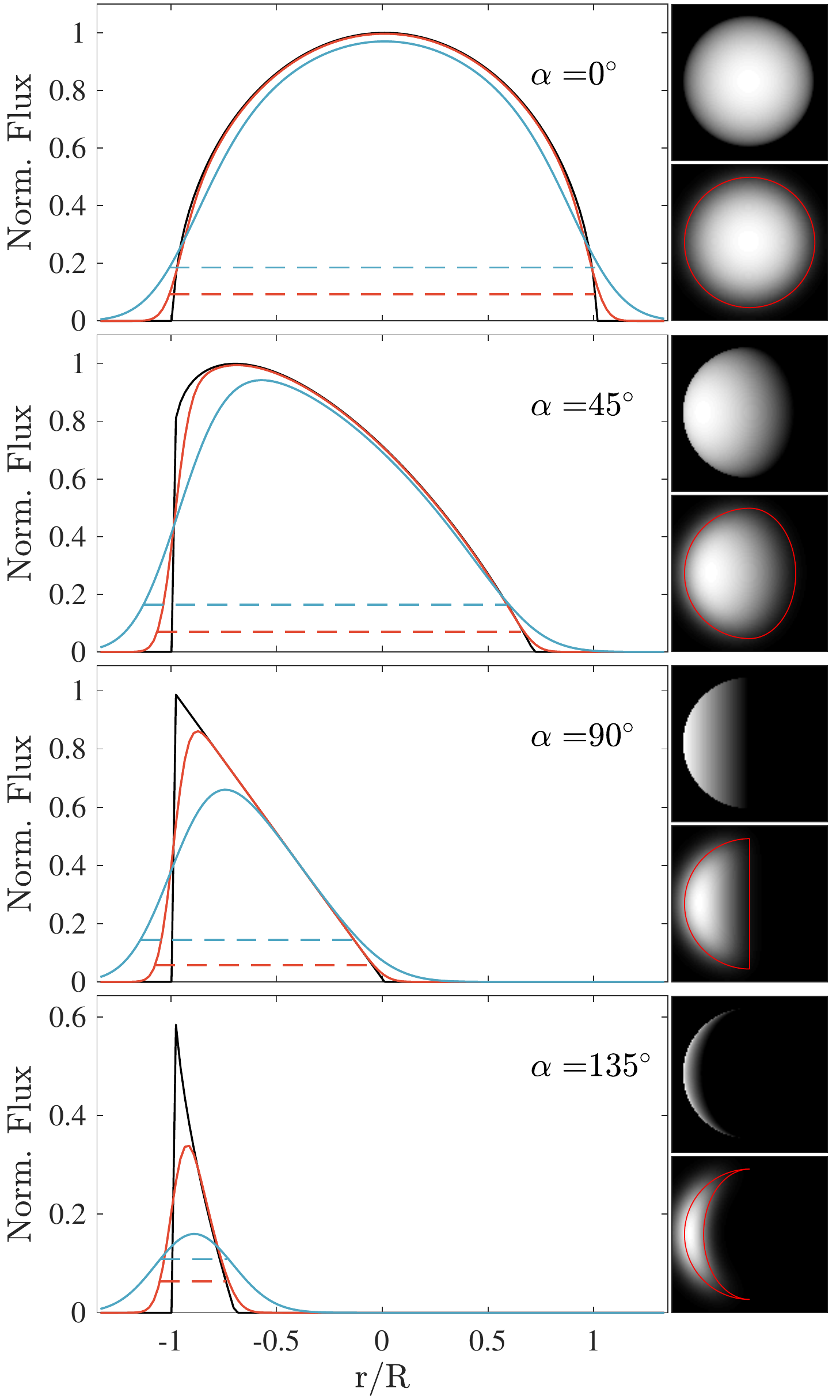}  
\caption{Simulation of a Lambert sphere of 15 arcsec of diameter seen with an atmospheric seeing of 0, 1 and 3 arcsec respectively (black, red, blue curves), and phase angles $\alpha=[0, 45, 90, 135]^\circ$. Lambert spheres with no seeing degradation are normalized with respect to the intensity at the sub-solar point. The length of the horizontal dashed lines are equal to the actual size of the planet without seeing degradation. Their ordinates correspond to the photometric threshold applied to match the observed planet with the expected size along the equator. The right frames represent the Lambert sphere without (top) and with a 3-arcsec seeing degradation (bottom). The red line indicate the edges of the sphere without seeing degradation.}
\label{fig_seing_position_cut}
\end{figure}


We simulate a Lambert sphere seen at different phase angles and degrade it with typical seeing conditions ranging from 0.5 to 4 arcsec. A four-arcsec seeing can sound quite large but is not uncommon for daytime observations, as it often happens for Venus (e.g., Gaulme et al., in prep.). In Fig. \ref{fig_seing_position_cut}, we present the cut of a Lambert sphere along the equator as well as the same observed with atmospheric seeing of 1 and 3 arcsec. Phase angles range from 0 to 135$^\circ$. At zero phase angle, atmospheric seeing impacts only the apparent diameter of the planet because the blurring effect is symmetrical around the edge. Then, as the phase angle increases, the planet seems to get shifted towards the Sun's direction. 
This is due to the fact that the limb edge is steeper than the terminator, making the seeing-caused blurring stronger on limb side than on the terminator. In Fig. \ref{fig_seing_position_cut}, we also represent two examples of images, without and with seeing, which show that the planet seems to drift away from its actual position.

\begin{figure}[t!]
\center
\includegraphics[width=9cm]{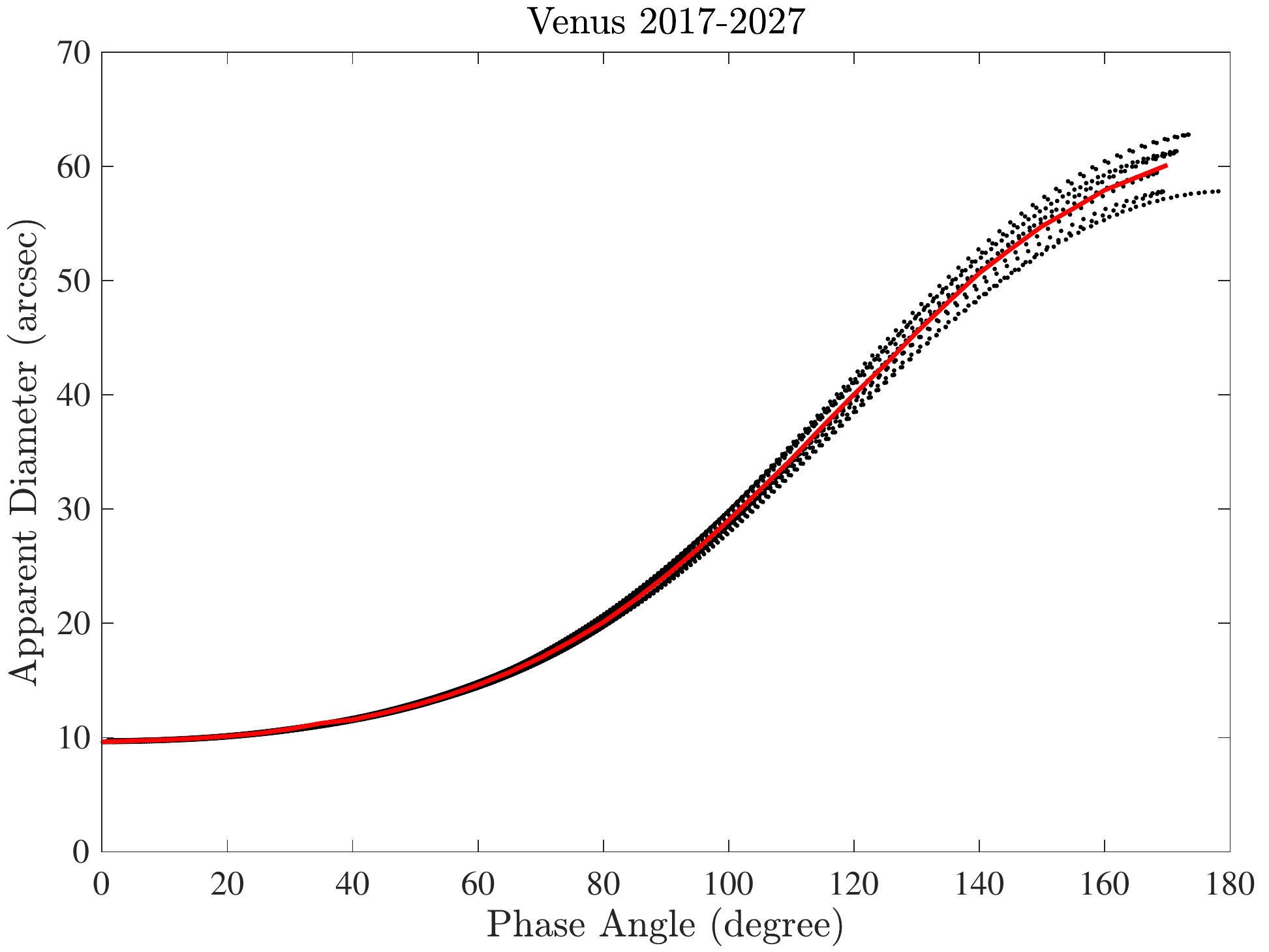}  
\includegraphics[width=9cm]{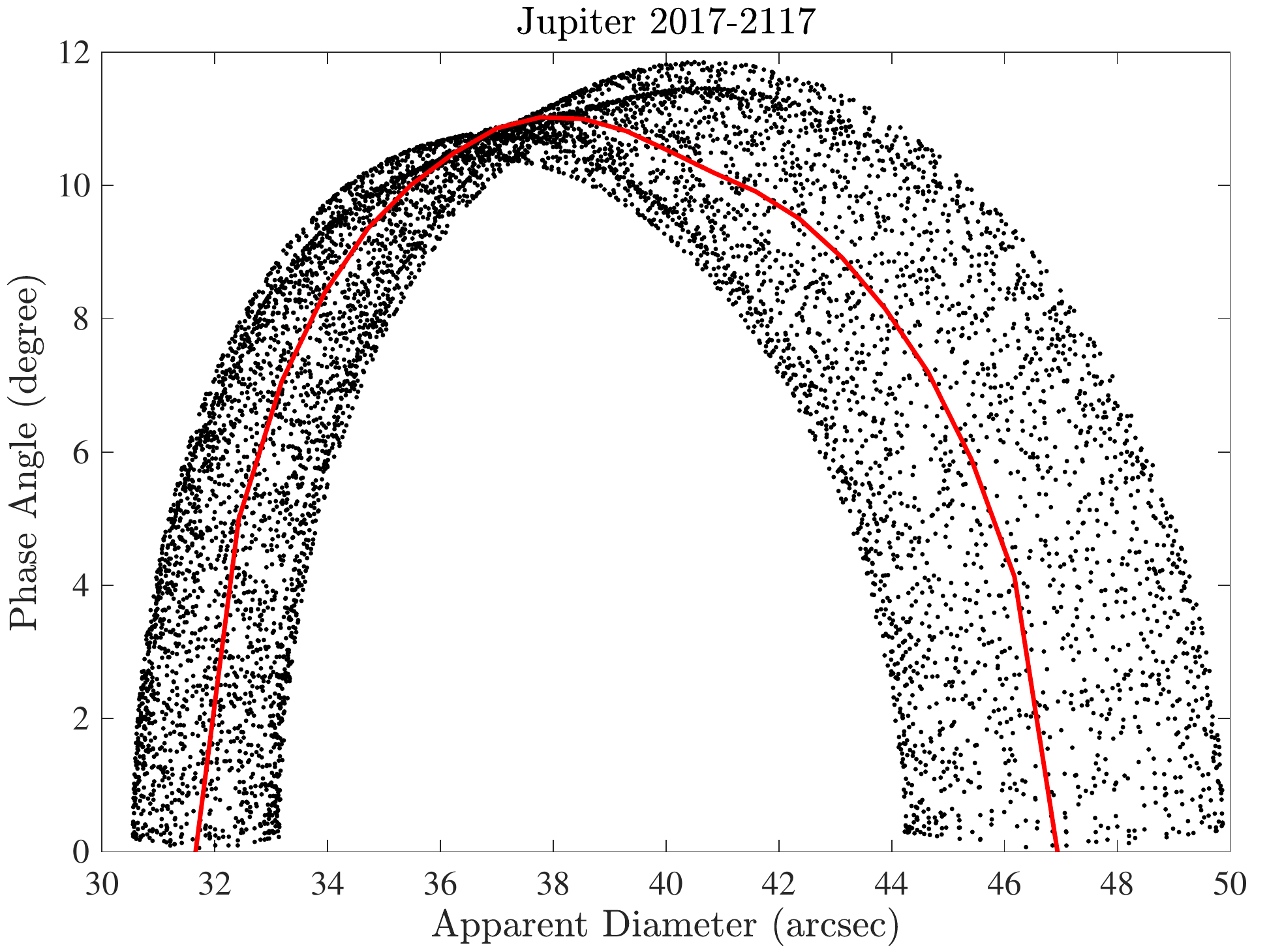}  
\caption{Top panel: apparent diameter of Venus (arcsec) as function of phase angle (degrees), from June 2017 to June 2027 (black dots). Bottom panel: same for Jupiter, with axis reversed with respect to Venus, and a longer time span (2017-2117) to average enough orbits.} For both panels, the plain red curves indicate moving averages of the black dots, which were then used to associate an (two for Jupiter) angular diameter to a given phase angle in Fig. \ref{fig_position_shift}.
\label{fig_venus_jupiter_phase_diam}
\end{figure}

To quantify how much seeing alters the apparent position of a planet, we simulate a Lambert sphere in the conditions Venus and Jupiter are observed. For this, we download the ephemeris of the two planets from the NASA \textit{Horizons} web interface (https://ssd.jpl.nasa.gov/horizons.cgi) on long time periods to get an average relationship connecting the phase angle and apparent diameter (Fig. \ref{fig_venus_jupiter_phase_diam}). For Venus, we choose to present the results as function of the phase angle as it is a critical criterion when observing it. Indeed, winds are impossible to measure when a small fraction of the disk is visible, and very low or very large phase angle correspond to bad observing conditions near solar conjunctions. As regards Jupiter, for a given phase angle, two possible apparent diameters are possible as it is an outer planet. This causes issues to produce a clear plot, as two results should be presented at each value of the phase. This is why we choose to present Jupiter simulations as function of apparent diameter.

\begin{figure}[t!]
\center
\includegraphics[width=9cm]{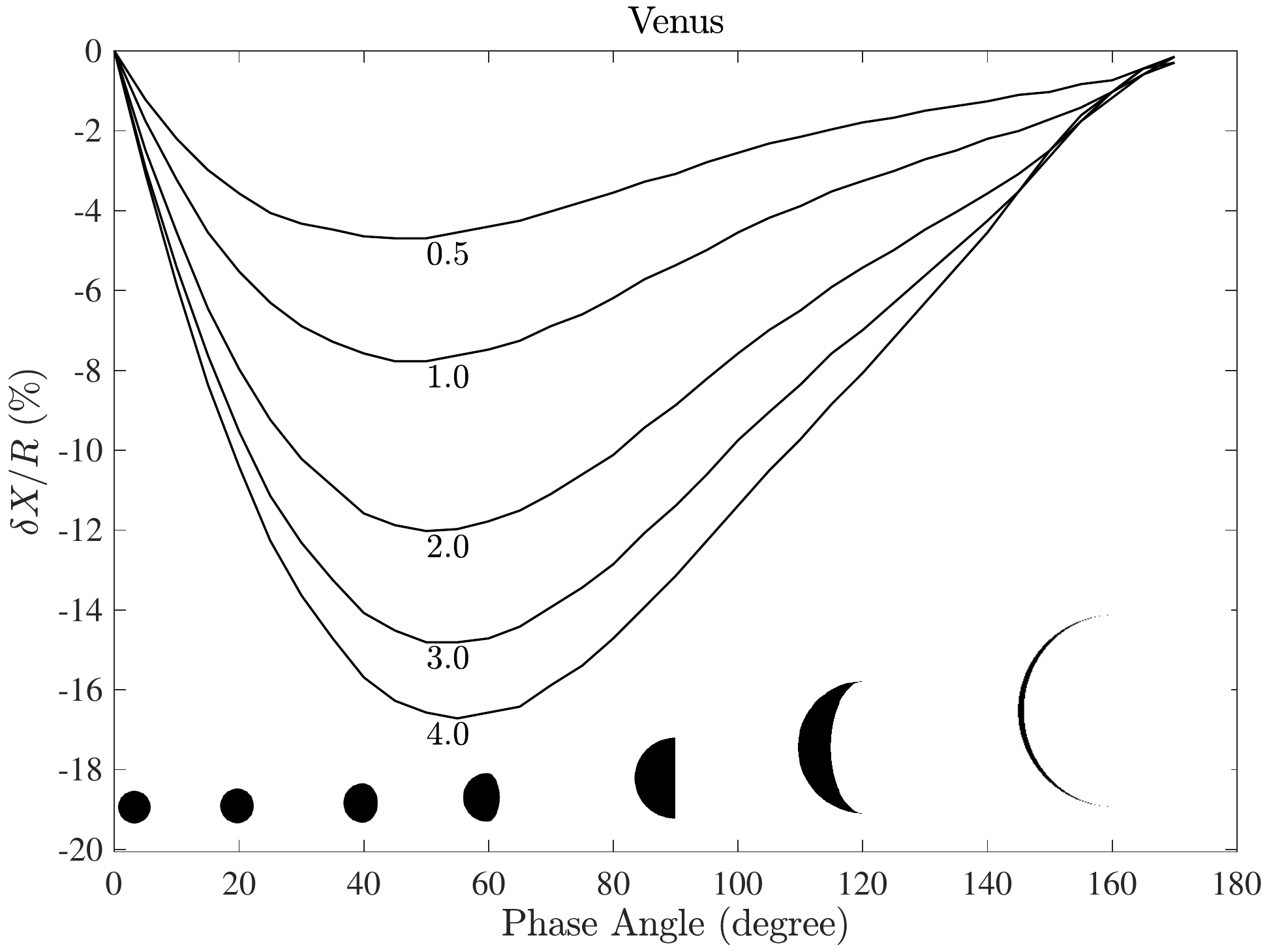}  
\includegraphics[width=9cm]{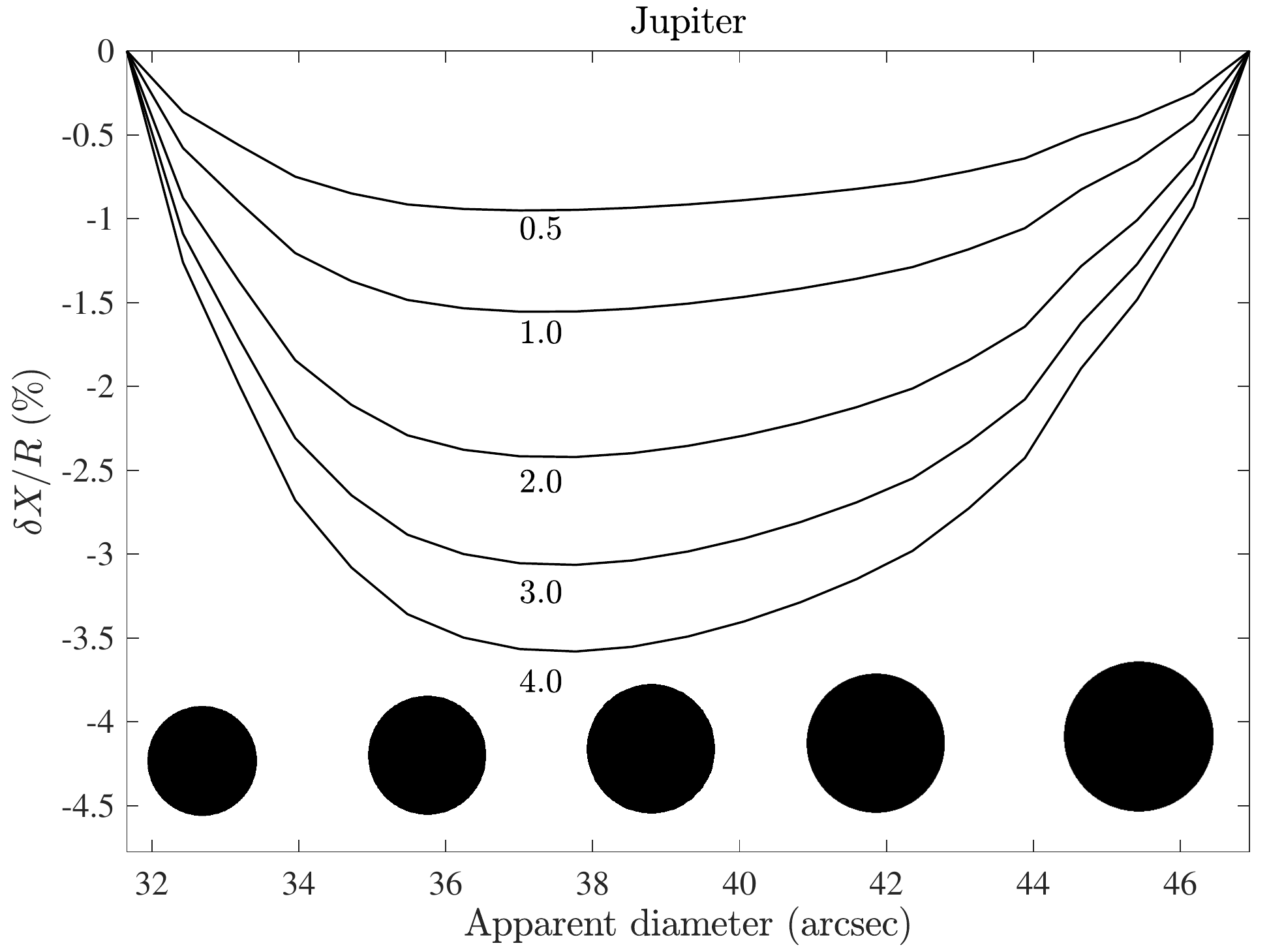}  
\caption{Apparent shift of a Lambert sphere in the conditions Venus (top) and Jupiter (bottom) are seen from Earth. For Venus case, the shift along the equator relatively to the planet's radius $\delta X/R$ is plotted as function of the phase angle (Sun-Venus-Earth). For Jupiter, $\delta X/R$ is plotted as function of the apparent diameter.  From top to bottom, the consecutive lines correspond to different seeing conditions, from 0.5 to 4 arcsec, as indicated near the minimum of each curve. The black shapes indicate the planet apparent sizes and phases. Note that proportions are conserved in between Venus and Jupiter panels. }
\label{fig_position_shift}
\end{figure}

We consider positive phase angles, which means the Sun is on the left side of the simulated planet. As indicated in the introduction of this section, we assume the pixel FOV and planet diameter to be known and determine the location of the planet from the photometric profile along the equator, as done in Fig. \ref{fig_seing_position_cut}. In Fig. \ref{fig_position_shift}, we display the apparent shift of ``Lambertian'' Venus and Jupiter along the $x$-axis from the position of the equator, expressed as a fraction of the planet's apparent radius. The common characteristic to both planets is that a zero phase angle leads to no apparent displacement of the planet, because the seeing symmetrically alters both sides. 

As regards Venus,  Fig. \ref{fig_venus_jupiter_phase_diam} firstly reminds us how small Venus is relatively to Jupiter if observations aim at  measuring winds, i.e. if more than half of the dayside is visible: its apparent diameter is less than 25 arcsec. Therefore, the seeing impacts it more than Jupiter. Then, we notice than beyond about $55^\circ$, the apparent displacement of the planet gets smaller to reach about zero at $180^\circ$. This is due to the fact that photometric profile near terminator is becoming steeper, as can be seen in Fig. \ref{fig_seing_position_cut} for $\alpha = 90$ and $135^\circ$. Next to $\alpha=180^\circ$ the dayside crescent is so thin that seeing degradation is practically symmetrical on limb and terminator edges, making the apparent displacement to be none. Overall, for Venus, the apparent shift of the planet is large at it reaches from 4 and 17\,\% of its radius within seeing conditions ranging from 0.5 to 4 arcsec. For a typical observation with 2-arcsec seeing and phase angle of 70 degrees the planet appears shifted of about 11\,\% of its radius. 

As regards Jupiter,  the maximum impact of seeing on positioning occurs at Jupiter's maximum phase angle, which is about $11^\circ$ for an apparent diameter of 38 arcsec. In such a case the apparent shift of the planet ranges from 0.8 to 3.6\,\%, which sounds almost negligible. However, Jupiter's equatorial velocity is about 12.5 km s$^{-1}$. Remembering that the Doppler signal of a reflected solar line is enhanced by a factor $(1 + \cos\alpha)$, i.e. almost 2 for $\alpha\leq11^\circ$,  the Doppler signal varies of about 50  km s$^{-1}$ along Jupiter's equator. A shift of 3\,\% along the equator is equivalent to a 1.5-km s$^{-1}$ radial velocity bias. It means that seeing impact on positioning needs to be considered too for a planet as Jupiter. We further develop some observation strategy to circumvent this positioning bias in Sect. \ref{sect_suchkinddadata}.

\subsection{Radial velocities}
\label{sect_seeing_rv}
As originally pointed out by \citet{Civeit_2005}, atmospheric seeing alters radial velocity measurements as it blends regions with non-uniform radial velocity and photometry. The resulting radial velocity is thus the convolution of the radial velocity signal with the photometric map of the considered object, including its degradation by seeing. The mean Doppler $\Delta V'$ measured in a given pixel $(x,y)$ on the detector can be expressed as: 
{\setlength{\mathindent}{0cm}
\begin{eqnarray} 
\Delta V'(x,y) & = & \displaystyle{\frac{(\Delta V\ F* P) (x,y)}{(F * P)(x,y)}} \label{eq_RV_convol_0}\\
& = & \displaystyle{\frac{\int\int \Delta V(u,w)\ F(u,w)\ P(x-u,y-w)\,\de u\, \de w}{\int\int F(u,w)\ P(x-u,y-w)\ \de u\,\de w}}
\label{eq_RV_convol}
\end{eqnarray}}
where $\Delta V$ and $F$ are the ``real'' radial-velocity and photometric maps \textit{prior} to seeing degradation, $P$ is the PSF of the atmospheric seeing, and the asterisk sign $*$ indicates the convolution product. This effect produces an artificial Doppler shift in opposition to the actual planetary rotation radial velocity. Indeed, because of the seeing, regions of larger photometric intensity and low Doppler signal contribute more to the Doppler map than the faint edges with high Doppler signal.  

\citet{Civeit_2005} originally studied this question for attempting to measure the rotation of Io, one of Jupiter's satellites. It was then used to measure the rotation of Saturn's moon Titan \citep{Luz_2005,Luz_2006}.  In both cases, the objects apparent sizes are smaller than the seeing PSF, which motivated the authors to take the seeing bias on radial velocities into account. The only visible Doppler-spectroscopic based observations of other planets regarded Venus, and none of the authors considered the seeing bias \citep{Widemann_2007, Widemann_2008,Gabsi_2008, Gaulme_2008, Machado_2012, Machado_2014, Machado_2017}.  The reason is likely that since the apparent diameter is significantly larger than the seeing PSF, most authors considered it to be negligible. In this section, we first show that such an assumption is erroneous, and second, that the \textit{Young} effect turns out to be much altered.

\begin{figure}[t!]
\center
\includegraphics[width=9cm]{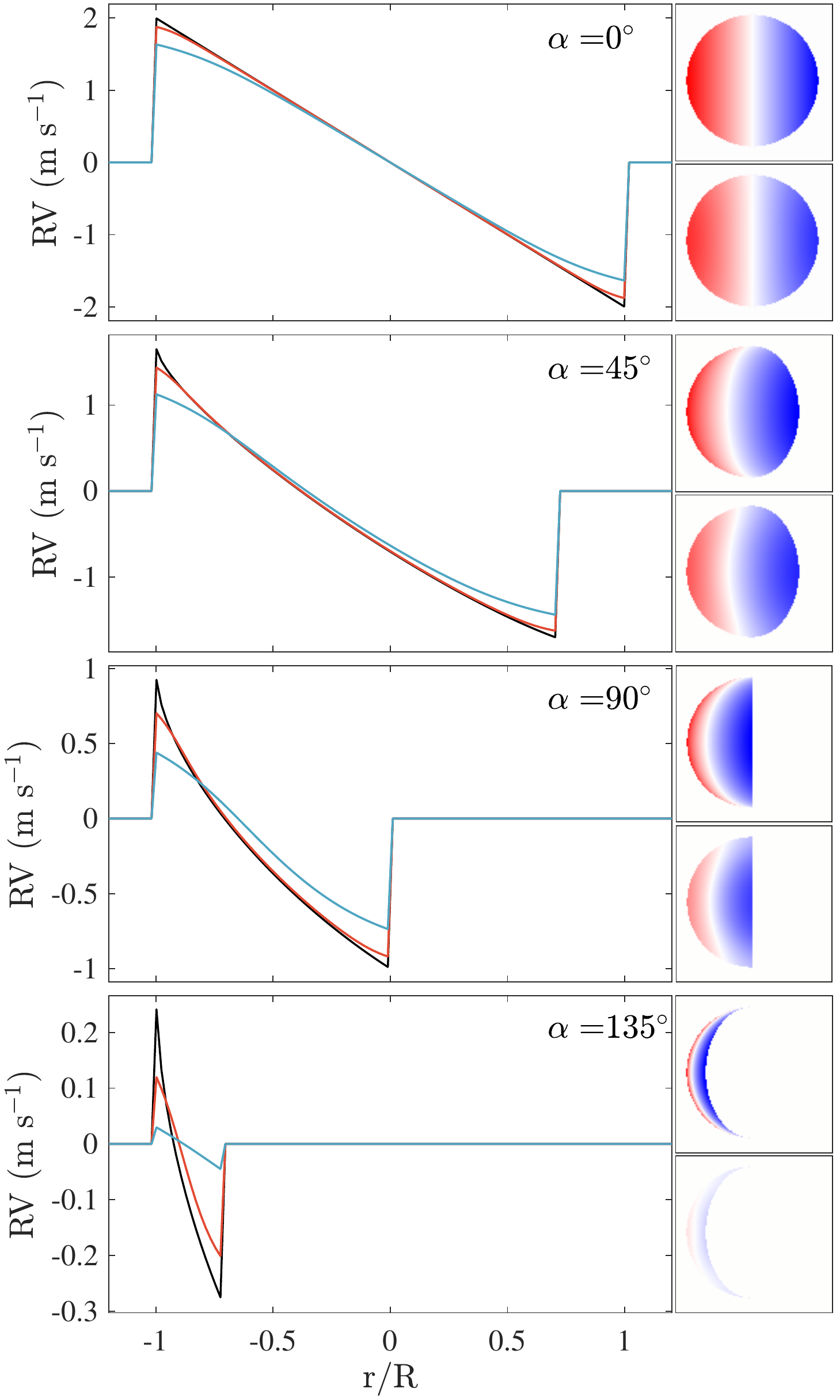}  
\caption{Radial velocities of a Lambert sphere of 15 arcsec of diameter, with a retrograde rotation of 1 m s$^{-1}$ seen with an atmospheric seeing of 0, 1 and 3 arcsec respectively (black, red, blue curves), and phase angles $\alpha=[0, 45, 90, 135]^\circ$. We assume that radial velocities are measured on solar reflected lines (Fraunhofer), which explains why radial velocities can be larger than $\pm 1$ m s$^{-1}$. Left panels are cuts along the equator and right frames are maps. Top maps are the true radial velocity signal, and bottom are the same degraded by a 3-arcsec atmospheric seeing.
}
\label{fig_RV_seeing_cut}
\end{figure}

For a zonal circulation, the atmospheric seeing reduces the amplitude of radial velocities because the regions where the Doppler shift is maximum display lower photometry. In Fig. \ref{fig_RV_seeing_cut}, we present simulations of a Lambert sphere that rotates as a solid-body, with 1-m s$^{-1}$ amplitude at equator,  15-arcsec apparent diameter, and phase angles (0, 45, 90, 135$^\circ$)  as in Fig. \ref{fig_seing_position_cut}. We assume radial velocities to be measured on solar Fraunhofer lines, so the Doppler shift is enhanced by $(1 + \cos\alpha)$. Radial velocities along the equator are compared to what is measured by assuming a 1- and 3-arcsec seeing. At zero phase angle, a 1-arcsec seeing reduces the measured radial velocity of about 6\,\% at the edge, whereas a 3-arcsec seeing reduces it of about 18\,\%. The larger is the phase angle, the larger is the bias on radial velocities. 

\begin{figure}[t!]
\center
\includegraphics[width=9cm]{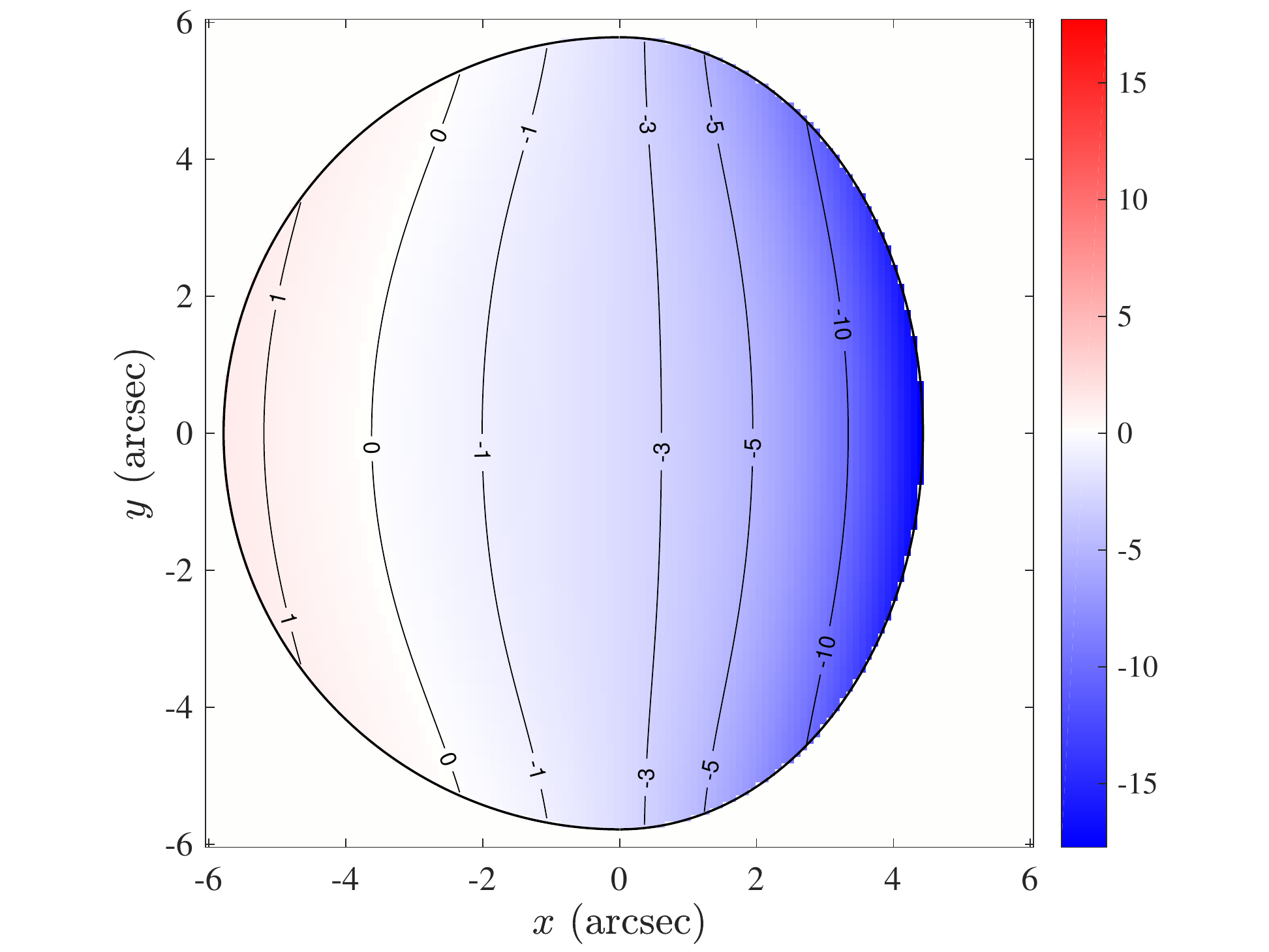}  
\includegraphics[width=9cm]{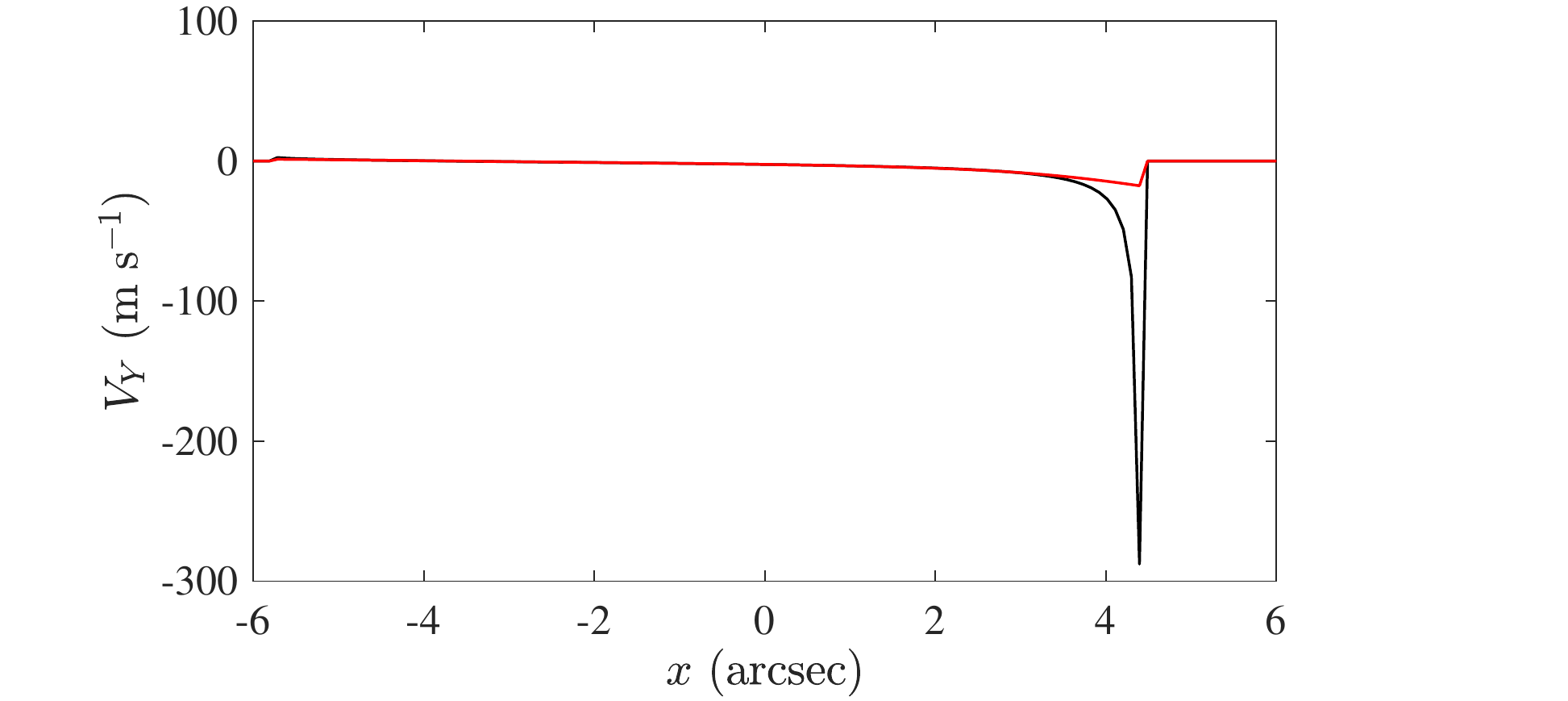}  
\caption{Top: map of the \textit{Young} effect on Venus as seen from the Earth when the atmospheric seeing's PSF is 2 arcsec, the phase angle (Sun-Venus-Earth) 40$^\circ$ and the apparent diameter 11.6 arcsec. Bottom: cut of the apparent \textit{Young} radial velocity $V\ind{Y}$ along the equator in the same conditions as above, with (red) and without (black) blurring by atmospheric seeing.}
\label{fig_young_simu_seeing}
\end{figure}

As regards the \textit{Young} effect, since its amplitude varies very steeply near the terminator, it is obviously much altered by atmospheric seeing. In Fig. \ref{fig_young_simu_seeing}, we present a map of the \textit{Young} effect in the case of Venus, in the same conditions as in Fig. \ref{fig_young_simu_sphere} ($\alpha=40^\circ$, $D\ind{app} = 11.6$ arcsec), as it appears with 2-arcsec atmospheric seeing conditions. As expected, where photometric variations are slow, the same order of magnitude is preserved, while towards the morning terminator, the maximum amplitude is reduced to 17 m s$^{-1}$ instead of about 2,000 m s$^{-1}$. This tells that if the \textit{Young} effect should not be neglected when searching for winds of about 100 m s$^{-1}$, as its amplitude can be over 10 m s$^{-1}$ in the early morning area, it also never reaches values as large as the solar rotation velocity, even with infinitely small pixels.

\subsection{Combined effect of seeing photometry and radial velocities}
In the previous two sections, we show that atmospheric turbulence: 1) generates an apparent shift of a planet seen with a non-zero phase angle; 2) biases radial velocities, especially close to the edges, where the photometric gradient is steep\footnote{Note that in the absence of a photometric gradient, the seeing would simply smooth the radial velocity map. Because of spatial photometric variability, such as limb darkening, the atmospheric seeing introduces a systematic bias in radial velocity measurements of any extended object.}. In this paragraph, we comment whether one or both of these effects lead to over- or underestimating the amplitude of a global atmospheric circulation pattern, especially in the zonal case.

\begin{figure}[t!]
\includegraphics[width=9cm]{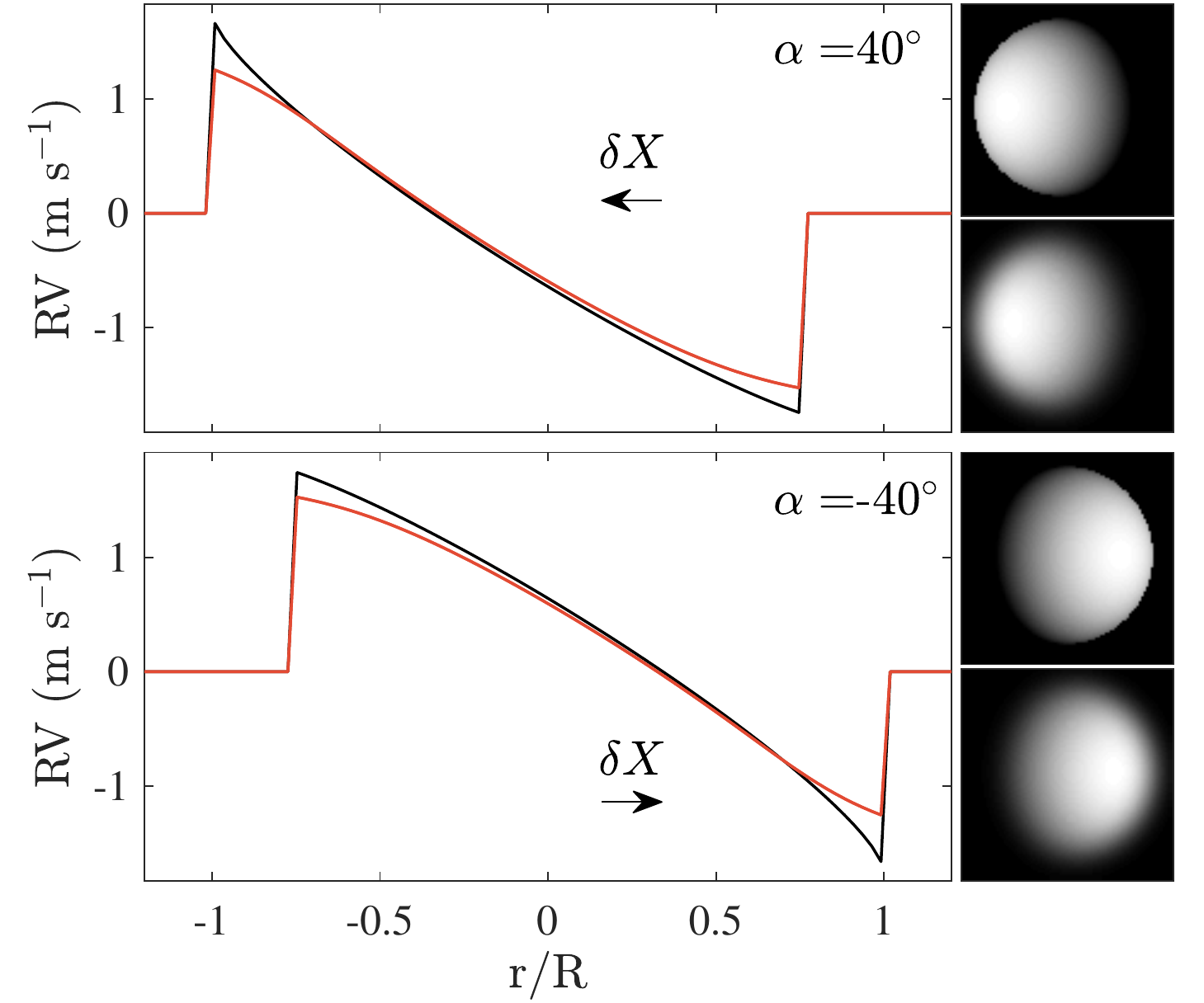}
\caption{Left panels: cut of the radial velocity map of a Lambert sphere rotating in the retrograde direction (solid-body rotation of 1 m s$^{-1}$ at equator), observed with a phase angle of $\pm40^\circ$. Radial velocities are measured on solar reflected lines. The width of the PSF used to degrade the image and radial velocity map is 35\,\% of the planetary radius, which is equivalent to a 2-arcsec atmospheric seeing for Venus when it is observed at $\pm40^\circ$ (see Fig. \ref{fig_venus_jupiter_phase_diam}). The term $\delta X$ refers to the apparent displacement of the planet on the sky caused by the atmospheric seeing.}
\label{fig_seeing_combined_effect}
\end{figure}

The effect of a PSF on estimating a wind circulation is non-linear and impossible to fully predict. Nonetheless, we identify different possible scenarios that depend on the observing configuration. Let us consider the case of an inner planet, as Venus, where the atmospheric circulation is purely zonal. Let us also neglect the Young effect, given it is almost canceled by the PSF.  
In the case of a correct positioning of the spectrometer on the planet -- apparent shift is taken into account -- the reduction of the radial velocity field towards the edges leads to systematically underestimating the zonal wind (Fig. \ref{fig_seeing_combined_effect}).

To the contrary, when the positioning on the planet is erroneous -- apparent shift is \textit{not} taken into account --, the effect of the seeing can lead to either case. We remind that the apparent displacement of the planet caused by atmospheric seeing is towards the Sun, wherever the phase angle is positive or negative (Fig. \ref{fig_seeing_combined_effect}). If the spectroscopic measurements are away from the limb and terminator regions, as it is often the case for the Venus observations led by Widemann and Machado, the apparent displacement of the planet leads to overestimating the wind amplitude because the radial velocity profile gets monotonically steeper towards the limb. In the case spectroscopic measurements cover the whole planet, the wind estimation is the result of the competing phenomena: underestimation caused by reduced radial velocities close to the edges, and overestimation by apparent shift. The outcome is unpredictable a priori. In any case, at some point, if the atmospheric seeing is very large -- enough to alter the whole radial-velocity profile, as with the 3-arcsec case in Fig. \ref{fig_RV_seeing_cut} -- underestimating the wind amplitude  dominates.

Interestingly, the above comments could explain in part why zonal-wind measurements of Venus by \citet{Machado_2012, Machado_2014, Machado_2017} provide very consistent results -- both internally and with respect to cloud tracking --, with average  zonal winds estimated in between 117 and 123 m s$^{-1}$, whereas \citet{Widemann_2007,Gabsi_2008} report much lower values ($\approx 75$ m s$^{-1}$). The former results were obtained either with the CFHT or the VLT telescopes which are located on sites where the seeing is very good (maximum seeing reported is 1.4 arcsec in \citealt{Machado_2012}). To the contrary, both papers \citet{Widemann_2007} and \citet{Gabsi_2008} report bad observing conditions at the Observatoire de Haute Provence, which could explain why they obtain such low wind estimates. 

\subsection{How to deal with such kind of data}
\label{sect_suchkinddadata}
Except for space-borne observations, atmospheric seeing alters radial-velocity measurements. Adaptive optics could considerably reduce the seeing effect, but it has never been used so far for this type of observations, because adaptive optics systems are relatively rare on high-resolution spectrometric devices. Adaptive optics is planned as part of the second stage of the JOVIAL/JIVE project \citep{Goncalves_2016}. In this section, we establish a data analysis protocol to optimize the reliability of such kind of observations. 

The first step is to know where ``we are'' on the planet. Having a partial or full image of the planet in parallel to spectroscopic data is fundamental. It could be a guiding camera image as in Widemann- or Machado-led papers, or a slice of planet as in \citet[][and in prep]{Gaulme_2008}. The image is the result of the apparent diameter of the planet, the telescope optics and the atmospheric turbulence. Let suppose the photometric limb-darkening profile of the planet to be known, one could retrieve an estimate of both the atmospheric seeing and the pixel FOV, by fitting a photometric profile degraded by a PSF. This is the approach used by Gaulme et al. (in prep.) for Venus, as the observations were taken mostly during daytime so there was no way to monitor the atmospheric seeing. Different levels of refinements are possible for this approach. Firstly, the pixel FOV can be calibrated by observing apparently close stars (e.g., wide multiple-stars systems), which reduces the number of free parameters. Secondly, the limb darkening law of the planet can be very simple (e.g., Lambertian), quite refined by including improved description of light scattering, or even based on space-borne data \citep[e.g.,][]{Mayorga_2016}. From Gaulme et al. (in prep.) and Gon\c{c}alves et al. (submitted) about Venus and Jupiter respectively, a Lambert approximation is usually enough. Indeed, Venus' smooth appearance in the mid-visible and small apparent diameter (5 to 10 times a typical seeing value) does not justify models that are too complex to be distinguished from a Lambert sphere given the observation quality. As for Jupiter, its apparent diameter is usually much larger than the seeing, and it is observed at low phase angle, where the seeing effect is radially symmetrical. The ideal situation involves the presence of a Differential Image Motion Monitor \citep[DIMM,][]{Vernin_1995} at the same observatory, which points at the same location on the sky (atmospheric seeing can substantially vary according to azimuth and altitude).

\begin{figure}[t!]
\center
\includegraphics[width=8cm]{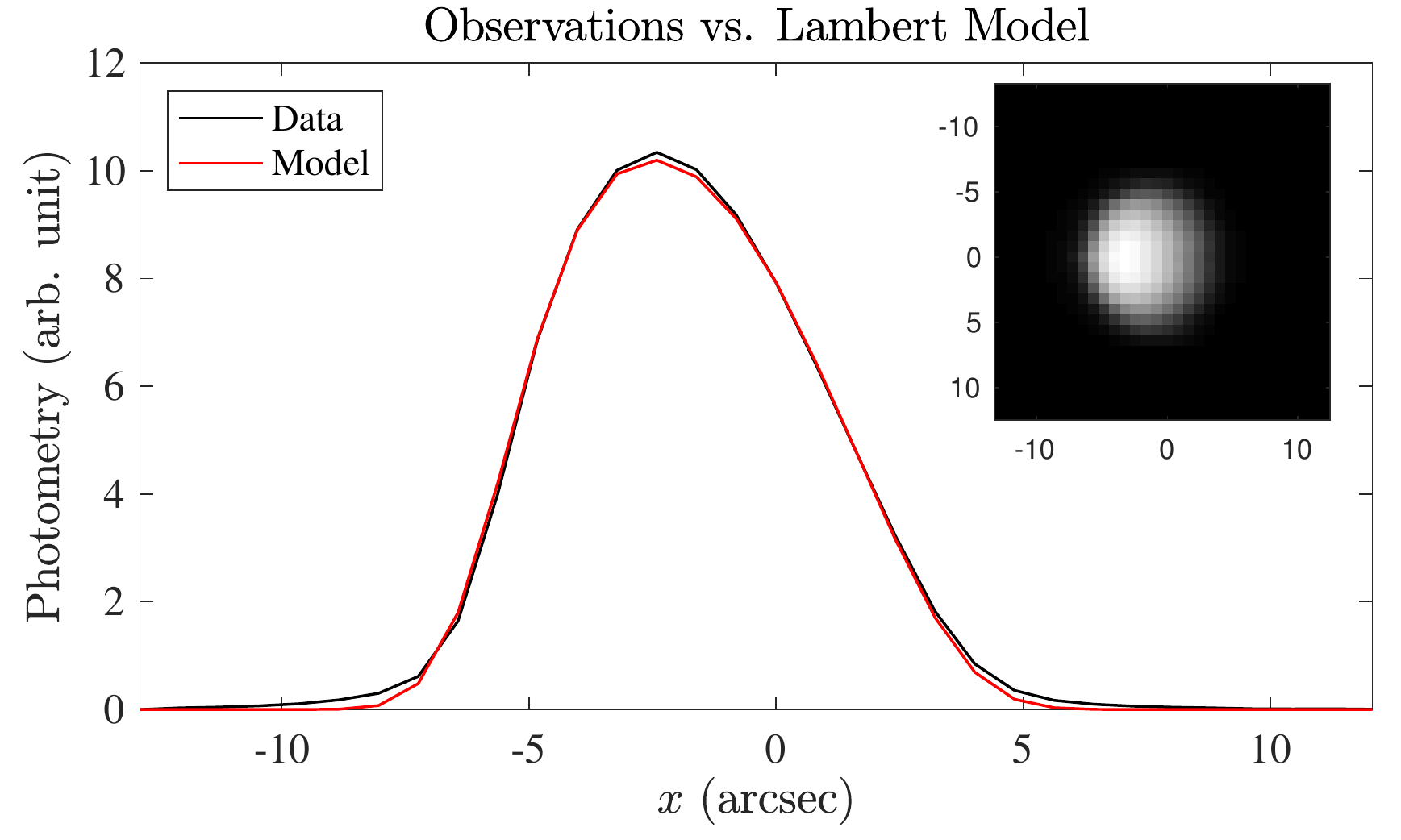}  
\includegraphics[width=8cm]{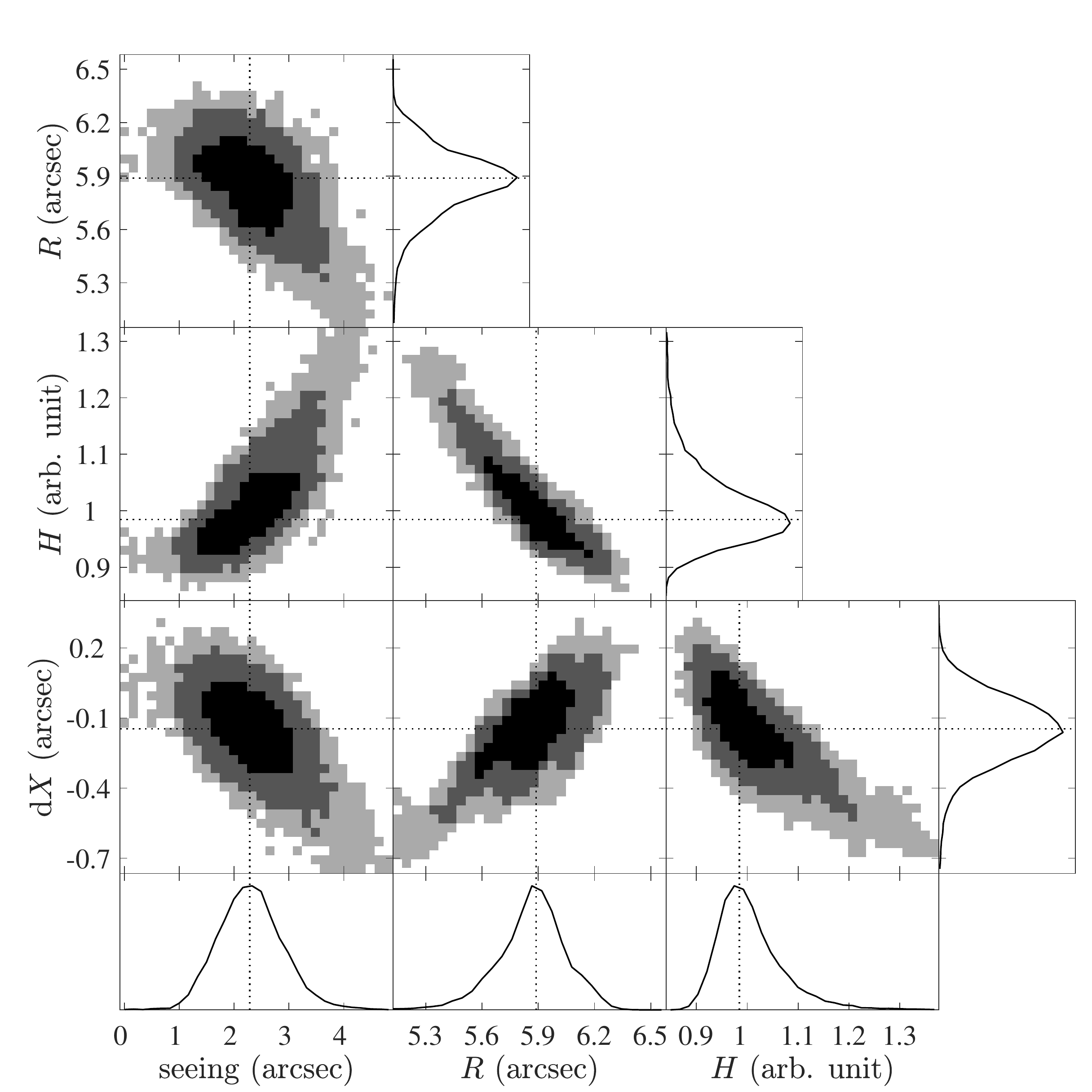}
\caption{Modeling Venus photometric profile with a Lambert law. Top panel, sub-frame: image of Venus obtained with the THEMIS solar telescope (Gaulme et al., in prep.). Top panel, main frame: the black line is the projection of the enclosed image along the $x$-axis, and the red line is the best-fit profile of it. Bottom panel: the above fitting was obtained by maximizing the likelihood of a Lambert-sphere model  to the THEMIS observations, with an MCMC algorithm.  Each sub-panel represents the two-parameter joint posterior distributions of all free parameters. The 68.3\,\%, 95.5\,\%, and 99.7\,\% confidence regions are denoted by three different gray levels. Probability Density Functions (PSFs) are plotted at the bottom and the right of the panels. The dotted lines mark the maximum probability value of the PDF of each parameter. After $200,000$ iterations, the fitted values are: $2.3_{-0.6}^{+0.5}$ arcsec for the seeing, $5.89_{-0.23}^{+0.12}$ arcsec for the apparent radius, $0.98_{-0.05}^{+0.07}$ photometric amplitude (here normalized to the maximum value), and $-0.15_{-0.16}^{+0.11}$ arcsec for the planet's decentering with respect to the center of the detector. The red curve on the upper panel is the result of the fitted values.}
\label{fig_model_lambert_themis}
\end{figure}

In Fig. \ref{fig_model_lambert_themis}, we show a photometric profile of Venus obtained at the Teide observatory together with a fit of it by a Lambert limb-darkening law. This profile is part of a larger dataset which will be presented in Gaulme et al. (in prep.). Fitting was performed with a by a dedicated Monte-Carlo and Markov Chain (MCMC) routine, based on the Metropolis-Hasting algorithm and parallel tempering, and Bayesian inference. Free parameters are the apparent diameter $R$, the seeing, as well as the maximum photometric flux $H$ and the position $\delta X$ of the center of the planet along the $x$-axis with respect to the center of the detector. Note that for computational reasons,  we fitted the apparent radius of the planet on the detector instead of the pixel FOV. Obviously, the actual apparent size of the planet in the sky is perfectly known from ephemeris databases. Note also that we need to fit $H$ even after normalizing the photometry to the maximum value (as in Fig. \ref{fig_model_lambert_themis}) because the maximum of intensity onto the detector's discretized grid is not the actual maximum of the photometric profile. The error-bars on the free parameters were obtained from their posterior density function, sampled by the MCMC. This illustrates that even with a simple Lambertian model, the fitted model is very close to the measurements, except at the very edge. 

\section{Inversion of Doppler spectroscopic observations}
\label{sect_model}
\subsection{Projection factors}
\label{sect_proj_fac}
In this section, we assume the above-described issues about location caused by the seeing are taken into account. We assume that whatever the technique (pinhole, long-slit, atomic cell, or Fourier transform spectrometry) we have a partial or full radial velocity map of the planet. 

Spectral lines may be Doppler-shifted by planetary atmospheres for four possible causes: zonal ``z'', meridional ``m'', vertical ``v'', and/or subsolar-to-antisolar  ``SA'' motions.
To invert radial velocity maps, we need the projection factors which transform any motion into a radial velocity field:
\begin{equation}
\Delta V\ind{i}(\lambda,\phi)   =  f\ind{i}(\lambda,\phi)\ V\ind{i} (\lambda,\phi) 
\end{equation}
where the subscript ``i'' is either ``z'', ``m'', ``v'', or ``SA'', and the coordinates $(\lambda,\phi)$ indicate the latitude and longitude in the planet system of coordinates. Details about projection factors are presented in Appendix \ref{app_proj_fac}.
In the case of reflected solar lines, the projection coefficient are expressed as follows:
\begin{equation}
f\ind{i} = -\vec{v\ind{i}}.\vec{O} - \vec{v\ind{i}}.\vec{S} 
\label{eq_proj_fac}
\end{equation}
where the vectors $\vec{O}$ and $\vec{S}$ point towards the observer and the Sun, and $\vec{v\ind{i}}$ is the vector describing the considered wind pattern of unitary amplitude, all expressed in the planetary frame (see Appendix \ref{app_proj_fac}).
Note that in the case we work with planetary emission or absorption lines, instead of solar reflected lines, the radial velocity with respect to the Sun is zero, i.e.,  $f\ind{i} = -\vec{v\ind{i}}.\vec{O}$.

\subsection{Deprojection versus global modeling}
\label{sect_invert}
At any given time, a radial velocity map is possibly the result of zonal, meridional, vertical or subsolar-to-antisolar motions. 
It is obviously impossible to retrieve a value of each circulation component at each point of the planet from a single map. In other words, some approximations are needed to get quantitative information on atmospheric circulation. The choice of approximation depends on the type of dominant circulation in the considered atmosphere, as well as the level of detail we seek for. Global properties of atmospheric circulation can be obtained by globally fitting an ensemble of measurements, whereas deprojection -- dividing an observed map by a projection factor -- can be performed either when only one circulation component is significant, or on small scales in case of high-resolution observations from a spacecraft, for example. We now review what is appropriate for the planets of the solar system.

Venus is the planet that has been studied the most with high-resolution Doppler spectroscopy in the visible domain. We know the circulation is dominated by a strong zonal wind, and a weaker equator-to-pole meridional circulation \citep[e.g.,][]{Machado_2017}. In such a case, it is not accurate to simply divide a radial-velocity map by $f\ind{z}$ to estimate the zonal wind at any latitude and longitude, as it completely neglects the meridional circulation. It is more appropriate to assume a circulation model composed of a zonal and a meridional component, as the vertical flow is likely to be negligible on global scale. Then, different hypothesis can be tested and compared in terms of model likelihood. For example, a zonal component can be assumed as cylindrical (uniform speed with latitude), solid-body (uniform angular rotation), or more complex, as assuming an individual amplitude coefficients per latitude. 
Note that subsolar-to-antisolar winds were measured in Venus mesosphere \citep[e.g.,][]{Lellouch_2008}, above 90 km. A possibly weaker return branch (antisolar-to-subsolar) deeper could possibly affect atmospheric dynamics measurements at cloud-top level where visible observations probe ($\approx 70$ km).
 Like Venus, Titan has an atmospheric super-rotation and similar modeling could be used. So far, its wind circulation was measured by  \citet{Luz_2005} in the visible, \citet{Kostiuk_2001,Kostiuk_2005,Kostiuk_2006,Kostiuk_2010} at 10-$\mu$m, \citet{Moreno_2005} in the mm range. These were ground-based observations before or in support of the Huygens Probe measurements, and only zonal circulation was considered.

The case of Jupiter and Saturn are very different as these two planets are dominated by significant differential rotation: once removed the rotation of their interiors, their atmospheres are structured in bands and zones that rotate in opposite directions.  Regarding meridional flows, because the band/zone alternate structure are small versions of Hadley cells, we do not expect global meridional circulation. Some intra band (or zone) meridional flows may exist but they are likely not measurable from Earth because of the relatively low spatial resolution\footnote{Note that if high-resolution planetary images are obtained from the ground by amateur astronomers, it is only with lucky imaging techniques, i.e., with exposure time less than 0.1 second. Radial velocity measurements are the results of stacking hours of observations, composed of individual exposures of at least 30 sec, and are thus limited by seeing.}. If local vertical motion of several tens of m s$^{-1}$ are expected in some jovian planet thunderstorms \citep{Hueso_2002}, vertical motions on large scales are a priori negligible with respect to zonal circulation. 
 In such a configuration, where only zonal circulation is considered, deprojection can be applied. Each map can be divided by $f\ind{z}(\lambda,\phi)$ (see details in next sub-section) and a value of the zonal wind is obtained in each point a any time. In practice, no observation is able to provide a reliable estimate from a single map: several hours are needed to be stacked to make the signal get out of the noise (see Gon\c{c}alves et al., submitted). A complete planisphere of zonal wind can be obtained by averaging data from several consecutive nights. This implicitly assumes that the zonal circulation does not evolve over a few days, and that there are no variations as function of local time either. This assumption is reasonable for Jupiter and Saturn, thanks to their fast rotation ($\sim10$ hours), which allows for complete longitude coverage in two to three consecutive nights. 
At last, Uranus and Neptune are not easy targets to be observed from the ground, as their apparent diameters are barely larger than typical seeing conditions (about 3.7 and 2.2 arcsec respectively). No such band/zone structure is visible, even though jets are observed at several latitudes for both planets.

\subsection{Implementing wind inversion}
\label{sect_invert}
As it is not possible to obtain all the wind components from a single image, we describe the best possible method for interpreting a radial-velocity map, by taking into account the peculiarities of the atmospheric dynamics of the considered planet. 
On one hand, if we have good scientific reasons to expect the wind structure to be composed of simple functions of local coordinates, we can express the model as a function a limited number of parameters, and try to estimate these parameters. That way, it is possible to monitor possible temporal evolution based on observations taken at different times. 
On the other hand, for a one-component wind circulation, the best approach consists of deprojecting the radial-velocity map because the result is not limited by model assumptions.

\textbf{Deprojection.} Let us consider the case of a one-component global circulation pattern, where zonal circulation dominates. In principle, the recipe to deproject a radial velocity map into a zonal map  is simple, as it consists of dividing the measured radial velocity map $\Delta V\ind{obs}$ by the geometrical projection factor $f\ind{z}$:
\begin{equation}
V\ind{z} = \frac{\Delta V\ind{obs}}{f\ind{z}},
\end{equation}
which was done, for example, by \citet{Machado_2012} with observations of Venus with the UVES spectrometer of the Very Large Telescope (VLT).
However, this is absolutely true only in absence of seeing degradation. In practice, the actual radial velocity map is described by Eq. (\ref{eq_RV_convol_0}). In the simple case of pure zonal circulation, and by expressing $\Delta V=f\ind{z} V\ind{z}$, Eq. (\ref{eq_RV_convol_0}) rewrites as: 
\begin{equation}
\Delta V\ind{obs} = \displaystyle{\frac{(f\ind{z}\ V\ind{z}\ F)* P}{F * P}} 
\end{equation}
where we dropped the $(x,y)$ dependence to ease the reading. By assuming the angular rotation to be constant within the size of the PSF, we can take the term $V\ind{z}$ out of the parenthesis and get:
\begin{equation}
\Delta V\ind{obs} \approx V\ind{z}\ \displaystyle{\frac{(f\ind{z}\ F)* P}{F * P}};
\end{equation}
then:
\begin{equation}
V\ind{z} \approx \frac{\Delta V\ind{obs}}{f\ind{z}}\ \mathcal{C}\qquad \mbox{where }\qquad \mathcal{C} \equiv f\ind{z} \frac{F * P}{(f\ind{z}\ F)* P}.
\end{equation}
The introduction of the $\mathcal{C}$ factor is the way to take into account the biases introduced by the atmospheric seeing, otherwise velocity maps are biased. Our ability to deproject is a matter of knowing the three parameters  $f\ind{z}$, $F$, and P. The term $F$ is the photometric model, i.e. the actual image deconvolved from seeing degradation. The term P is the PSF and is estimated in the data processing steps. 

The retrieved velocity map $V\ind{z}$ tends to infinite where the projection factor is close to 0. Therefore, a threshold needs to be applied to keep only values of the radial velocity that are reasonable. This is done by computing the noise map associated to the velocity map:
\begin{equation}
\sigma\ind{z} = \frac{\sigma\ind{\Delta V}}{|f\ind{z}|}\ \mathcal{C}
\end{equation}
The above equation shows that deprojection is not a good way to retrieve the properties of wind circulation because uncertainties tend to infinite where radial velocities tend to zero. 
We can circumvent this limitation whenever the data consists of a cube of images taken at different orientations of the planet. Within the assumption that the wind only depends on local coordinates of the planet and can be regarded as constant during all the observations, we can average several deprojected maps with different orientations and avoid to have regions with high levels of noise. To do so, individual deprojected maps are converted into spherical coordinates, and images with different rotation angle, i.e., looking at different longitudes, are averaged with a weighted mean, to discard data with larger noise level from the final map. Estimation of Jovian zonal wind maps obtained in this way will be presented in a future paper by Gon\c{c}alves et al. (in prep.).


\textbf{Global modeling.} When atmospheric circulation needs to be described as the sum of several components (e.g., zonal and meridional), a global atmospheric model can be fitted on the whole map at once. The first step consists of minimizing $\chi^2$, the sum of the quadratic difference of the observations relatively to the model:
\begin{equation}
\chi^2  = \frac{1}{n}\sum_n{\left(\Delta V\ind{obs} - a\ind{z}\ \Delta V\ind{mod,z} - a\ind{m}\ \Delta V\ind{mod,m}  - \kappa\right)^2}, 
\label{eq_chi2_noweight}
\end{equation}
where ``obs'' refers to the observed radial velocity map, ``mod'' to the model, ``z'' to zonal, and ``m'' to meridional;  $n$ is the number of pixels on the planetary disk, $a_z$ and $a_m$ are the amplitudes of radial velocity components, and $\kappa$ is a global offset in case of drift of the spectrometer. 
To take into account the atmospheric seeing effect, the models need to be priorly convolved with the PSF in order to minimize the $\chi^2$.

We just considered a very simple model with only three parameters, but the principle is the same for more complex models, as long as they linearly depend on the free parameters, in the form:
\begin{equation}
\Delta V\ind{obs}=H \alpha + N,
\label{eq_linear_LS}
\end{equation}
where $\alpha$ is the vector containing the $m$ parameters of the model to be estimated, and $n$ is the number of measurements, where $m<n$. The matrix $H$ is of rank $m$ with a $m\times n$ size. The random noise $N$ has independent components and null mean. An unbiased estimate of the parameters is obtained by minimizing the least square. Since the convolution by the seeing's PSF is a linear operation, this approach is still valid even when taking it into account. The convolution can be included in the model. If the noise is not uniform across the image -- it is certainly the case when photon noise dominates as flux is not uniform -- a weighting of the least square by the noise can be applied before minimization.  The vector of estimate parameters $\hat\alpha$ is retrieved by computing:
\begin{equation}
\hat\alpha = (H^tH)^{-1}\ H^t\ \Delta V_{obs},
\label{eq_estimate_LS}
\end{equation}
or in case of weighted least square:
\begin{equation}
\hat\alpha = (H^tWH)^{-1}\ H^tW\ \Delta V_{obs}
\label{eq_weight_estimate_LS}
\end{equation}
where the weighting function $W$ is a diagonal matrix where the diagonal terms are the inverse values of the noise of the data. 
The estimate $\hat\alpha$ of the parameter vector $\alpha$ is truly unbiased, however, as the sensitivity of Doppler signal to the different components is null at some locations, some caution has to be taken to define the model that can effectively be recovered from the data. For instance, a model with a non-zero meridional component at the equator is not only unlikely for physical reasons, but also impossible to estimate for a mathematical point of view. In practice, we should take care that the Gramian matrix $H^tH$ is not close to be singular, because it would result in a strong increase of the uncertainties on the estimated parameters. As the same procedure provides an estimate of these uncertainties, it is possible to check the goodness of the fit. A poor normalized value of $\chi^2$ indicates an inadequate model.

\section{Conclusion and prospects}
Despite the Doppler effect has been known for over 175 years, its application to planetary atmospheres in the visible has remained limited because of difficult implementation. Significant efforts in that matter have been led in the past decade, on one hand thanks to the ground-based follow-up organized to support the Venus Express mission \citep{Lellouch_Witasse_2008,Widemann_2007,Widemann_2008,Gaulme_2008,Gabsi_2008,Machado_2012,Machado_2014,Machado_2017}, on the other hand for the ground-based projects SYMPA and JOVIAL/JIVE dedicated to Jupiter seismic observations based on with visible Doppler imaging \citep{Schmider_2007,Gaulme_2008b,Gaulme_2011,Goncalves_2016}. As it is somehow a new domain for planetary observations, the purpose of this paper was to consolidate the theory for rigorous Doppler spectro-imaging observations. 

We first revised and modified the theoretical prediction of the artificial rotation caused by the Sun's rotation,  which was originally proposed and incompletely expressed by \citet{Young_1975}. We revised the analytical approach, and produced numerical simulations which refine the estimated amplitude of the phenomenon by including the Sun's limb darkening and differential rotation. We also explored the impact of atmospheric seeing on planetary Doppler observations. It arises that for planets that are seen at non-zero phase angles, the convolution of the actual planetary image with the seeing PSF shifts it on the detector, which leads to significant biases when associating latitude and longitude reference frames to images of a planet. In addition, we studied the impact of atmospheric seeing on radial velocity measurements, which was pointed out by \citet{Civeit_2005} for almost point-source objects (Titan, Io), and we showed it is significant as well for larger objects such as Venus or Jupiter. In the last section, we propose a roadmap for interpreting radial velocity maps, by including the \textit{Young} and atmospheric seeing effects. We expressed the projection factors or any zonal, meridional, vertical, or subsolar-to-antisolar motion, and presented two possible approaches to retrieve the wind circulation: direct deprojection or global modeling of radial velocity maps. 

This paper is the first of a series of three, the other two being in the submission process. Gaulme et al. (in prep.) present observations of Venus atmospheric dynamics obtained in 2009 with the MTR long-slit spectrometer of the THEMIS solar telescope located in Tenerife (Spain). Gon\c{c}alves et al. (submitted) presents observations of Jupiter with the JOVIAL/JIVE Doppler spectro-imager \citep{Goncalves_2016}, which was specially designed for studying atmospheric dynamics of the planets of the solar system as well as performing seismology of Jupiter and Saturn. They present average zonal wind profiles of Jupiter obtained in 2015 and 2016 at the Calern observatory (France), and compare them with simultaneous Hubble-space-telescope cloud tracking measurements \citep{Johnson_2018}. 
\section*{Acknowledgments} 
P. Gaulme was supported in part by the German space agency (Deutsches Zentrum f\"ur Luft- und Raumfahrt) under PLATO data grant 50OO1501. The JOVIAL project is supported by the Agence Nationale de la Recherche under the contract number ANR-15-CE31-0014-01. I. Gon\c{c}alves's PhD is granted by Observatoire de la C\^ote d'Azur and the JIVE in NM project (NASA EPSCoR grant \#NNX14AN67A). THEMIS is  UPS of the CNRS funded by INSU. The authors thank E. Lellouch for his very constructive review of the paper as the referee. P. Gaulme wishes to precise that he did most of the writing while listening to the 1969 Velvet Underground Live.


\bibliographystyle{aa} 
\bibliography{bibi_venus} 

\begin{appendix}

\section{Latitude and longitude of a 3D oblate spheroid}
\label{sect_3D_ell}
\begin{figure}[t!]
\center
\includegraphics[width=8cm]{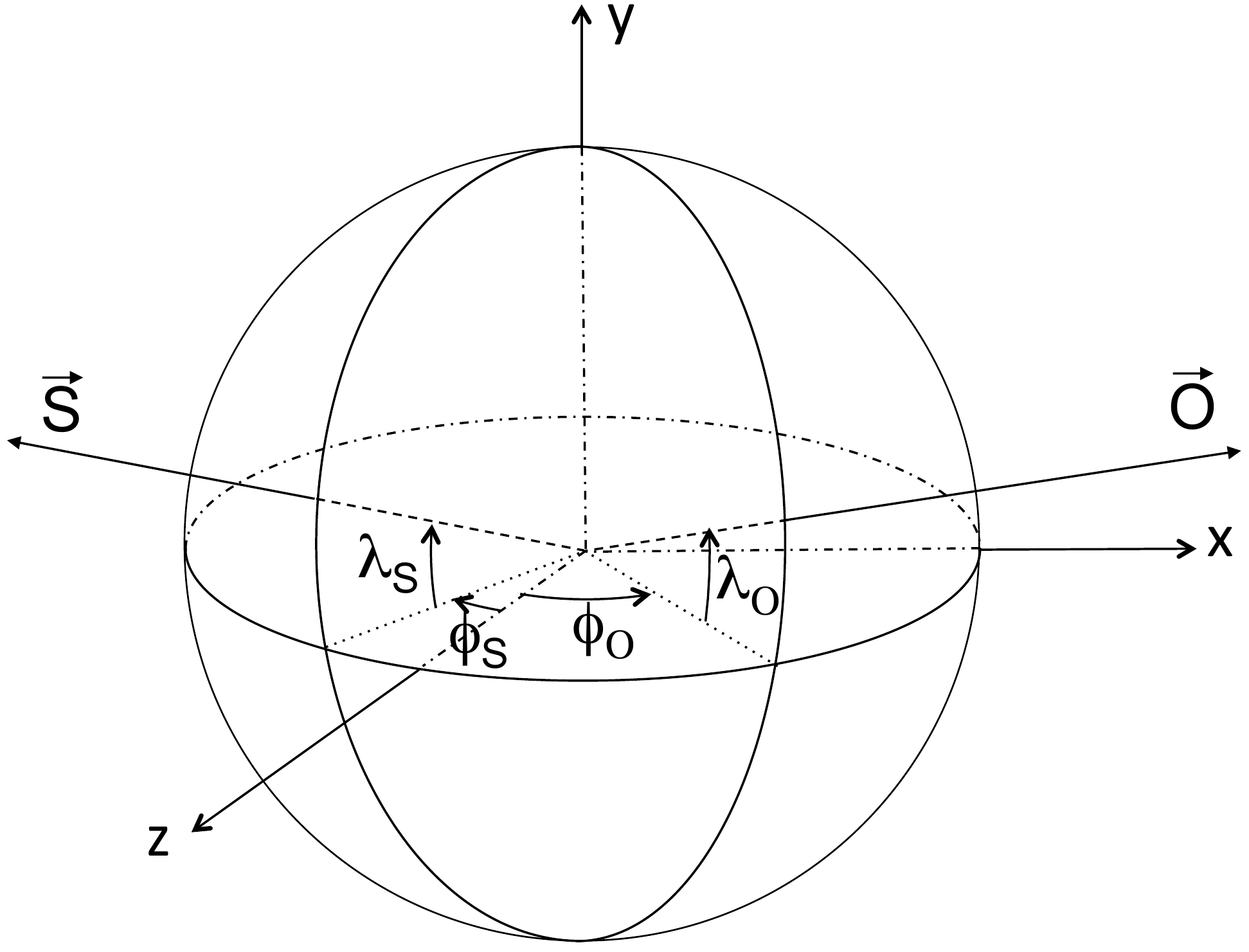}
\caption{
Planetary-centered cartesian system of reference $(x,y,z)$. The vectors $\vec{O}$ and $\vec{S}$ point towards the observer and the Sun, respectively. The angles $(\lambda\ind{O},\phi\ind{O})$ and  $(\lambda\ind{S},\phi\ind{S})$ are the latitudes and longitudes of the sub-observer and sub-solar points on the sphere. The $y$-axis corresponds with the planetary spin axis.
}
\label{fig_3D_frame}
\end{figure}

We consider the planet-centered system of coordinates $(x,y,z)$, defined such as the $y$ axis points at North (Fig. \ref{fig_3D_frame}). This frame is useful for avoiding confusion when dealing with data where an image of a planet is defined as function of the $(x,y)$ axis of a detector (e.g., the simulated Venus in Fig. \ref{fig_young_simu_sphere}). 
Rotating planets or stars can be approximated with an oblate spheroid, i.e., axes along $x$ and $z$ are the same.
For an oblate spheroid, the latitude $\lambda$ and longitude $\phi$ on the planet are simply derived from the planetary-centered Cartesian frame:
\begin{eqnarray}
\lambda &=& \cos^{-1} \sqrt{\frac{x^2 + z^2}{\rho^2}} \\
\phi        &=& \cos^{-1} \left(\frac{-x}{\rho \cos \lambda}\right) \frac{z}{|z|}.
\end{eqnarray}
In the case of a sphere $\rho = R$; for an oblate spheroid $\rho$ expresses as: 
\begin{equation}
\rho = \sqrt{R\ind{eq}^2 + \left(1 - \frac{1}{\mu^2}\right) y^2},
\end{equation}
where $R\ind{eq}$ is the equatorial radius and $\mu$ the oblateness factor defined as $R\ind{pole} = \mu R\ind{eq}$. 

\section{A Lambert sphere illuminated by an extended source}
\label{sect_lambert}
A Lambert sphere assumes that incoming light is isotropically scattered by its surface. A light coming from a given direction reaches the surface with an intensity that is proportional to the cosine of the incidence angle $\gamma$, defined as the angle between the incoming light direction and the normal to the surface. The intensity of a Lambert sphere is thus:
\begin{equation}
I(\gamma) = I_0 \cos\gamma,
\end{equation}
As illustrated in Fig. \ref{fig_solar_frame_photo}, the intensity received from the Sun in each point of a Lambert sphere is the solar photometry, which can be described by a limb-darkening law, modulated by the variation of incidence angle across the solar surface:
\begin{equation}
I(\gamma) = \frac{I_0}{\pi R^2} \int_{-R}^{R} \int_{-\sqrt{R^2 - y^2}}^{\sqrt{R^2 - y^2}}  I\ind{LD}(x,y) \cos(\gamma-y)\ \mathrm{d}x\ \mathrm{d}y
\label{eq_integral_intensity}
\end{equation}

\section{Projection factors}
\label{app_proj_fac}
The radial velocity map of a planet in the case of reflected solar-lines is the result of the relative motion at the surface of the planet with respect to both the Sun and the observer:
\begin{equation}
\Delta V = -\vec{V}.\vec{O} - \vec{V}.\vec{S} 
\end{equation}
where $\vec{V}$ is the velocity vector in the planet system of coordinates, $\vec{O}$ is the unitary vector pointing at the observer, and $\vec{S}$ is the unitary vector pointing at the Sun. The negative sign arises from the usual convention that a redshift corresponds with a positive radial velocity.  

Let us consider a sphere, for which $(\lambda\ind{O},\phi\ind{O})$ and $(\lambda\ind{S},\phi\ind{S})$ are the coordinates of the sub-observer and sub-solar points in the planetary frame (Fig. \ref{fig_3D_frame}). The vectors $\vec{O}$ and $\vec{S}$ are expressed as:
\begin{equation}
\vec{O} = 
\begin{pmatrix}
 \cos\lambda\ind{O} \sin\phi\ind{O}\\
  \sin\phi\ind{O}\\
 \cos\lambda\ind{O} \cos\phi\ind{O}
 \end{pmatrix}  
 \mbox{\qquad and \qquad }
 \vec{S} = 
 \begin{pmatrix}
 \cos\lambda\ind{S} \sin\phi\ind{S}\\
  \sin\phi\ind{S}\\
 \cos\lambda\ind{S} \cos\phi\ind{S}
 \end{pmatrix} 
\end{equation}

To express projection factors, we need to decompose a given circulation pattern (zonal, meridional, vertical, subsolar-to-antisolar) of unitary amplitude in the $(x,y,z)$ system of coordinates
A zonal motion is described as:
\begin{equation}
\vec{v}\ind{z} = \
\begin{pmatrix}
\cos\phi \\
 0\\
-\sin\phi
\end{pmatrix}  
\end{equation}
A meridional motion is described as:
\begin{equation}
\vec{v}\ind{m} = 
\begin{pmatrix}
- \cos\lambda \sin\phi\\
\cos\lambda\\
\sin\lambda \cos\phi
\end{pmatrix}  
\end{equation}
A vertical (radial) motion  is described as:
\begin{equation}
\vec{v}\ind{v} = 
\begin{pmatrix}
\cos\lambda \sin\phi\\
\sin\lambda\\
\cos\lambda \cos\phi
\end{pmatrix}  
\end{equation}
A sub-solar to anti-solar circulation is described as in the frame centered on the sub-solar point, in which $(\lambda\ind{S},\phi\ind{S}) = (0,0)$:
\begin{equation}
\vec{v}\ind{SA,\odot} = 
 \begin{pmatrix}
\cos\gamma \sin\theta\\
\cos\gamma \cos\theta\\
-\sin\gamma\\
\end{pmatrix}  
\end{equation}
In the planet system of coordinates, it is expressed as:
\begin{equation}
\vec{v}\ind{SA} =
  \begin{pmatrix}
     - \cos\lambda\ind{S} \sin\phi\ind{S} & -\sin\lambda\ind{S}  & \cos\lambda\ind{S} \cos\phi\ind{S}\\
    \cos\phi\ind{S}                                 & 0 &   -\sin\phi\ind{S}                                 \\
    -\sin\lambda\ind{S} \sin\phi\ind{S}  & \cos\lambda\ind{S}  & \sin\lambda\ind{S} \cos\phi\ind{S} 
 \end{pmatrix}
\ \vec{v}\ind{SA,\odot} 
\end{equation}

To compute the radial velocity field of a planet, we need to describe each component of the circulation model as follows:
\begin{eqnarray}
\vec{V}\ind{z}  &=& \cal{Z}(\lambda,\phi)\ \vec{v}\ind{z} \\
\vec{V}\ind{m}  &=& \cal{M}(\lambda,\phi)\ \vec{v}\ind{m} \\
\vec{V}\ind{v}  &=& \cal{V}(\lambda,\phi)\ \vec{v}\ind{v}\\
\vec{V}\ind{SA}  &=& \cal{S}(\gamma,\theta)\ \vec{v}\ind{SA}
\end{eqnarray}
where the functions $\cal{Z}$, $\cal{M}$, $\cal{V}$, and  $\cal{S}$ carry the information about the wind structure. For example, a solid-body circulation is $\cal{Z}(\lambda,\phi) = V\ind{eq}\cos\lambda$, where $V\ind{eq}$ is the velocity amplitude at equator. 

\end{appendix}


\end{document}